\documentstyle[epsf]{l-aa}  

\newcommand{\rf}{\par\noindent\hangindent 20pt}
\newcommand{\accrate}{$\dot{{\rm M}}_{{\rm acc}}$}
\newcommand{\massloss}{$\dot{{\rm M}}_{{\rm wind}}$}
\newcommand{\OI}{[OI]$\lambda$6300}
\newcommand{\LOI}{L$_{{\rm [OI]}}$}
\newcommand{\LHA}{L$_{{\rm H\alpha}}$}
\newcommand{\LSTAR}{L$_*$}
\newcommand{\LACC}{L$_{\rm acc}$}
\newcommand{\LIRE}{L$_{{\rm ire}}$}
\newcommand{\LPHOT}{L$_{{*, \rm nir}}$}

\newcommand{\be }{\begin{equation}}
\newcommand{\ee}{\end{equation}}
\newcommand{\kms}{kms$^{\rm -1}$}

\newcommand{\MS}{M$_{\odot}$}

\newcommand{\AAP}[2]{A\&A {#1}, #2}

\newcommand{\AAS}[2]{A\&AS {#1}, #2}

\newcommand{\AJ}[2]{AJ {#1}, #2}

\newcommand{\APJ}[2]{ApJ {#1}, #2}

\newcommand{\APJS}[2]{ApJS {#1}, #2}

\newcommand{\BAAS}[2]{BAAS {#1}, #2}

\newcommand{\MN}[2]{MNRAS {#1}, #2}

\newcommand{\PASP}[2]{PASP {#1}, #2}

\begin{document}
 
 \thesaurus{06   
        (08.13.2; 
        08.16.5; 
	08.03.4) 
       }
  \title{Wind diagnostics and correlations with the near-infrared
  excess in Herbig Ae/Be stars\thanks{Based on observations made at
  the La Palma Observatory, the Caltech Submillimeter Observatory, and
  the European Southern Observatory/Max Planck Institute 2.2m Telescope.}}
 
\author{M.\ Corcoran \inst{1,2} 
\and T.P.\ Ray \inst{1}}
      
\offprints{Myles Corcoran, E-mail: corcoran@discovery.saclay.cea.fr}
 
\institute{ Dublin Institute for Advanced Studies,
        5 Merrion Square, Dublin 2, Ireland
\and Service d'Astrophysique, Centre \'Etudes de Saclay,
 Gif-sur-Yvette 91101, France} 
 
\date{Received date ; accepted date}
 
  \maketitle
  \markboth{M.\ Corcoran \& T.P.\ Ray: Wind Correlations with the Near IR Excess}{}
  \begin{abstract}

Intermediate dispersion spectroscopic observations of 37 Herbig Ae/Be stars 
reveal that the equivalent widths of their [OI]$\lambda$6300 and H$\alpha$ 
emission lines, are related to their near-infrared colours in the same fashion 
as the T-Tauri stars. Such a correlation strongly supports the idea that the 
winds from Herbig Ae/Be stars arise in the same manner as those from T-Tauri 
stars, i.e.\ through accretion driven mass-loss. We also find that the 
[OI]$\lambda$6300 line luminosity correlates better with excess infrared 
luminosity than with stellar luminosities, again supporting the idea that 
Herbig Ae/Be winds are accretion driven. If one includes the lower mass 
analogues of the Herbig Ae/Be stars with forbidden line emission, i.e.\ the 
classical T-Tauri stars, the correlation between mass-loss rate and infrared 
excess spans 5 orders of 
magnitude in luminosity and a range of masses from 0.5${\rm M}_{\odot}$ to 
approximately 10${\rm M}_{\odot}$. Our observations therefore extend the 
findings of Cohen et al.\ (1989) and Cabrit et al.\ (1990) for low mass young 
stars and, taken in conjunction with other evidence (Corcoran \& Ray 1997a), 
strongly support the presence of circumstellar disks around 
intermediate mass stars with forbidden line emission. An implication of our 
findings is that the same outflow model must be applicable to these Herbig 
Ae/Be stars and the classical T Tauri stars.



   \end{abstract}
 
%

\section{Introduction}

Herbig Ae/Be stars (HAEBES) were first identified by Herbig (1960) as
possible higher mass analogues of the classical T-Tauri stars
(CTTS). Subsequent work has shown that the two classes of young stars
share many observational characteristics in common over and above
those one might expect on the basis of observational selection
criteria alone.  For example, like CTTS, the emission from a number of
HAEBES is intrinsically and variably polarized (Jain \& Bhatt 1995),
many have IR excesses (Hillenbrand et al.\ 1992), water masers (Palla
\& Prusti 1993) and evidence for accretion as seen in the redshifted
absorption components of their NaD lines (Sorelli et al.\ 
1996). Moreover most HAEBES possess the signatures of strong mass
loss, as indicated by P-Cygni line profiles (Finkenzeller \& Mundt
1984; P\'erez et al.\ 1992; Imhoff 1994), broad forbidden emission
lines (B\"ohm \& Catala 1994; Corcoran \& Ray 1997a) and by their
thermal radio continuum emission (Skinner et al.\ 1993).  Estimates of
their mass loss rates vary, and a number of different techniques have
been used (for details see Nisini et al.\ 1995), but typical values
are 10$^{-8}$ to 10$^{-6}$ \MS/yr.

The high level of activity of this group (for a general review of
HAEBES activity the reader is referred to Catala 1989) is perhaps
initially surprising if one attempts to understand it from a purely
stellar perspective. If its source is the star alone then one would
expect the mass loss to be driven by a sub-photospheric convection
zone. However, according to conventional theory (e.g.\ Gilliland 1986)
energy transport in HAEBES is purely radiative. How then can one solve
what B\"ohm \& Catala (1995) call the ``paradox'' of HAEBES activity?
Two approaches have been taken (B\"ohm \& Catala 1995): the first
assumes that there is an inherent source of energy, such as internal
rotation (Vigneron et al.\ 1990), which powers a stellar dynamo, while
the second attributes the activity to viscous disk accretion in a
manner similar to the CTTS (Hillenbrand et al.\ 1992). The viscous
accretion disk model might, at first sight, seem more plausible given
the observed IR excesses of most HAEBES (e.g.\ Hillenbrand et al.\
1992), however, alternative explanations for the IR excesses have
been put forward, based on the idea that HAEBES are surrounded by
quasi-spherical clouds of gas and dust (Berrilli et al.\ 1992;
Hartmann et al.\ 1993; Miroshnichenko et al.\ 1997). At the
same time, however, a number of HAEBES are associated with collimated
optical and molecular outflows (Mundt \& Ray 1994) and given, what
appears to be, the intimate connection between disks and outflows in
CTTS (see, for example, Edwards et al.\ 1993) this would lead one to
think that disks are present at least around a significant fraction of
the HAEBES population. Thus the question still remains whether the
standard disk model, developed to explain many of the observational
features of CTTS (e.g.\ Bertout et al.\ 1988), is applicable to the
higher mass HAEBES, or at least to those which, like the CTTS, have
forbidden line emission (see also Ghandour et al.\ 1994; Corcoran
\& Ray 1997a)? 

If accretion disks, perhaps in association with stellar magnetic
fields, provide the energy for driving HAEBES winds, one would expect
a correlation to exist between the accretion rate, \accrate, and the
wind mass-loss rate, \massloss.  Such a correlation is found for the
CTTS (Cabrit et al.\ 1990; Edwards et al.\ 1993), however, there are a
number of important points that must be borne in mind before one can
attempt a similar study for the HAEBES. Both \accrate\ and
\massloss\ are difficult to quantify and so appropriate substitutes
have to be found. In the case of the CTTS, Cabrit et al.\ (1990) and 
Edwards et al.\ (1993) used CTTS forbidden [OI] line emission as a measure of 
wind strength and their infrared excess as a measure of accretion rate. 
Such an approach for CTTS can be justified since there is clear evidence that 
the forbidden [OI] line emission originates in a wind (e.g.\ Edwards et 
al.\ 1993) and because the infrared excess in CTTS correlates with the veiling
continuum flux (see, for example, Hartigan et al.\ 1995), supporting the 
notion that the excess arises from viscous accretion. Can one use an identical 
strategy for HAEBES?

In the literature there has been some question as to whether the
forbidden line emission in HAEBES arises in the same manner as in CTTS
(see, for example, B\"ohm \& Catala 1994). A recent study by us
(Corcoran \& Ray 1997a) however shows clear parallels between the
forbidden line emission as observed in HAEBES and CTTS (see, for
example, Hartigan et al.\ 1995).  In particular double line profiles,
consisting of high and low velocity blueshifted components, are found
in a number of HAEBES although, like CTTS (Hartigan et al.\ 1995),
single low velocity blueshifted emission is by far the norm.  The
observation that the forbidden line emission of HAEBES is usually
blueshifted, as with the CTTS, strengthens the conclusion that, it is
formed in a stellar wind. Moreover the lack of any corresponding
redshifted components also argues that an obscuring disk is present
(Corcoran \& Ray 1997a). We will therefore take the approach in this
paper that {\em the strength of the forbidden line emission can be
used to gauge the strength of the wind from a HAEBES}. There is some 
question whether the same approach can be taken in the case of the 
hydrogen emission lines. For example, it was thought in the past
that the H$\alpha$ line could be used as a gauge of wind strength; current 
modeling of H$\alpha$ line profiles would suggest, however, that it is probably 
a {\em better measure of accretion} (Calvet 1997). In any event, we study 
both [OI] forbidden and H$\alpha$ line strengths here to see whether the same 
phenomenological relationships exists between these quantities as in the 
CTTS. Moreover we also look at how both quantities depend on our 
more `traditional' measure of accretion, again to test the analogy with CTTS.  

Turning now to the measurement of \accrate\ in HAEBES, this is
somewhat problematic for various reasons. Accretion disk theory would
suggest that we use either the excess UV emission, which may arise
from magnetic accretion columns close to the star, or that part of the
infrared excess which derives from viscous dissipation in the
circumstellar disk. The UV fluxes of these stars are poorly known
(although, indications of a far UV excess in a number of cases is
observed, Grady et al.\ 1993) so we are forced to consider the
infrared excess. The infrared excess, however, has two components
which are difficult to disentangle: the first is due to reprocessed
starlight in the disk and the second is due to heating by viscous
accretion.  For an optically thick, but geometrically thin, flat disk
up to one quarter of the luminosity of the star, depending on the
viewing angle, can be ``reprocessed'' (see, for example, Adams et al.\
1987 and Strom et al.\ 1993). Moreover if the disk is ``flared''
(Kenyon \& Hartmann 1987) even more light can appear as an ``excess''
although the contribution due to flaring is thought to be only
important at mid-infrared or longer wavelengths. In the case of the
CTTS samples (e.g.\ Hartigan et al.\ 1995), one is dealing with a
large number of stars of similar stellar photospheric luminosity but
with widely different accretion rates, thus despite the
``contaminating'' effects of reprocessed light, statistically one can
readily distinguish, for example, a correlation of the forbidden line
strength with the accretion luminosity. The situation with HAEBES is
not so clear-cut: HAEBES stellar luminosities can differ by factors of 
10$^{\rm 5}$ (Hillenbrand et al.\ 1992) and one is dealing
with statistical samples that are comparable in number to the CTTS
samples.  Thus a high degree of dependence on the stellar luminosity
is introduced since, as one might expect, in absolute terms both the
amount of reprocessed light and, to some extent, the accretion
luminosity (see, e.g.\ Hillenbrand et al.\ 1992) tend to scale with
the stellar luminosity.  This tendency for both the accretion and
reprocessed contributions to increase with stellar luminosity in
HAEBES, coupled with their large range in luminosities, gives rise to,
as we shall see, unavoidable correlations of the strength of the
forbidden line emission not only with the infrared excess, as per
CTTS, but with the stellar luminosity as well. A similar result was
found by Nisini et al. (1995) when they studied a small sample of 14 HAEBES, 
although in their case they erroneously (see above) used the intensity of 
near-infrared HI recombination lines as a measure of the wind strength. 
Obviously any correlation between the strength of the forbidden line emission
and the stellar luminosity, can be used to support the idea, expressed
by several authors (e.g.\ B\"ohm \& Catala 1994), that HAEBES winds
are stellar rather than accretion driven.  In the absence of a large
sample of HAEBES within a narrow stellar luminosity range,
statistically the question then becomes one of whether or not
forbidden line emission {\em is more tightly correlated with infrared
excess than with stellar luminosity.}

An alternative approach to circumvent any strong dependence on stellar
luminosity is, in a sense, to ``factor it out'' as much as possible.
In particular one can compare {\em the equivalent width} of forbidden
line emission, and for that matter Balmer emission, with {\em infrared
colours} since it is well known that the relative strength of the
forbidden line emission, for example, increases with increasing
redness in CTTS samples (Edwards et al.\ 1993). Moreover, Hartigan et
al.\ (1995) found a clear correlation between the ``veiling'' index
and K-L colour in CTTS, implying essentially a link between the ratio
of the accretion luminosity and the stellar luminosity, \LACC/\LSTAR ,
and near-infrared colour.  If the equivalent width of the forbidden
line emission in HAEBES is then {\em found to vary in the same way
with colour as with the CTTS}, then this would clearly point to
accretion as the driving force for HAEBES winds. We initially adopt
this promising approach here before considering any correlations of
our wind measures with stellar or excess infrared luminosities.

Finally, a relationship is also looked for between the strength of the
HAEBES winds, as measured by the [OI] line luminosity, and the amount
of circumstellar dust present, as indicated by mass estimates from mm
continuum measurements. We find, in contradiction to the result of
Nisini et al.\ (1995), who used a smaller sample, that there may be a
weak correlation between the luminosity of the [OI] line and the
amount of circumstellar matter. As will be described in the text,
however, one has to be cautious in interpreting this result.  

After outlining the details of our observations and data reduction
techniques in \S 2, we present the results of our study in \S 3 and
discuss its consequences in \S 4.

\section{Observations and data reduction}

Our data derive primarily from three observing runs in La Palma on the
Isaac Newton Telescope (INT) with the Intermediate Dispersion
Spectrograph (IDS) from 28 August -- 4 September 1991, 6--12 July 1993
and 14--20 December 1994. Further observations were taken in La Silla
with the ESO/MPI 2.2m Telescope using EFOSC-II from 23 December 1991
-- 1 January 1992. The instrumental details are listed in Table 1.

\begin{table}
\scriptsize
\begin {center}
\begin{tabular}{||l|c|c|c|c||}\hline
Spec.\ & Grating+CCD & Disp.\ & Res.\ & Range \\ 
& & (\AA\ pixel$^{-1}$) & (\kms) & (\AA) \\
\cline{1-5}
IDS &	R632V+GEC6	& 0.70	& 66  &	400 \\
&	R632V+EEV5	& 0.70	& 66  &	715 \\
&	R1200Y+GEC6	& 0.36	& 34  &	205 \\
&	R1200Y+EEV5	& 0.36	& 34  &	370 \\
&	R1200Y+TEK3	& 0.39	& 39  &	500 \\
&	H1800V+GEC6	& 0.22	& 21  &	120 \\
EFOSC 2 & Grism 9 & & & \\
 & +Thompson CCD& 1.11	& 105 & 1145 \\
\hline 
\end {tabular}
\vspace{1mm}
\caption{IDS and EFOSC 2 grating+CCD parameters. The resolution (Res.) of
the gratings and grism are calculated assuming 2 pixels for full
sampling. The pixel size of the Thompson CCD is 19$\mu$m, all other
CCDs have a pixel size of 22$\mu$m. Data taken from ING User Manual
VII and EFOSC 2 Operating Manual.}
\end {center}
\normalsize
\end{table}

	The combined dataset from the three La Palma observing runs
and the La Silla run form a database of H$\alpha$ and/or \OI\
observations of 57 Herbig Ae/Be stars. A subset was selected of this 
database of stars for which full photometric data was available in the 
literature (from the R to the N bands) and for which we had an exposure 
including the \OI\ region. A total of 37 stars met these two 
criteria and are listed in Tables 2 and 3. The excluded stars
include both those for which we were unable to obtain (due to time
constraints and/or weather problems) an exposure of the \OI\ region
and those with photometric coverage insufficient to provide the
necessary data for the calculation of the near infrared luminosity and
excess.

Data reduction was carried out using standard
IRAF\footnote{The IRAF software is distributed by the National Optical
Astronomy Observatories under contract with the National Science
Foundation} routines. Bias subtraction and flat-fielding corrections
were determined from zero second exposures and tungsten lamp exposures
respectively. The adjacent sky spectrum was subtracted from the object
spectra, and the dispersion solutions were determined from CuAr arc
exposures.

Photometric data for HAEBES in the infrared and millimeter
wavelengths were in most cases taken from the literature (Hillenbrand
et al.\ 1992; Hamann \& Persson 1992) although it was found necessary
to supplement these data with additional mm observations in the case
of some stars. These millimeter continuum data were obtained at the
Caltech Submillimeter Observatory (CSO) on Mauna Kea in April 1992.  A
$^3$He-cooled germanium bolometer served as our detector with a beam
size of 30$''$ (FWHM). Uranus and 3C\ 273 were used as calibration
standards. The CSO data was reduced as described in Beckwith \&
Sargent (1991).

\begin{table*}
\footnotesize
\begin{center}
\begin{tabular}{||l|c|c|c|c|c|c|c|c|c|c|c||} 
\hline
Name                       &  SpTy     &  Dist  &  $A_{\rm v}$&  Grp. &  W([OI])   &  W(H$\alpha$) &  H-K &  K-L     &  L-M     &  M-N     &  12-25$\mu$m\\
                           &           &  (pc)  &  (mag)   &        &  (\AA)       &  (\AA)    &  (mag)   &  (mag)   &  (mag)   &  (mag)   & (mag) \\
\cline {1-12}			      	       
LkH$\alpha$ 198 	   &  A5       &  600   &  2.5 	   &  II    &  1.37        &  40.50    &  1.43    &  2.08    &  0.28    &  2.95    &  2.71 \\
V376 Cas 		   &  F0       &  600   &  2.9 	   &  II    &  1.61        &  33.40    &  1.77    &  1.86    &  0.15    &  3.54    &  2.67 \\
BD+61 154 		   &  B8       &  650   &  2.1 	   &  I     &  0.18        &  44.10    &  1.05    &  1.30    &  0.20    &  2.08    &  1.79 \\
Elias 1\dag 		   &  A6       &  160   &  4.1 	   &  II    &  $\leq$0.02  &  20.60    &  0.98    &  1.00    &  0.40    &  3.14    &  2.80 \\
AB Aur 			   &  A0       &  160   &  0.4 	   &  I     &  0.11        &  26.30    &  0.67    &  1.08    &  0.38    &  2.52    &  2.14 \\
HK Ori 			   &  A5       &  460   &  1.2 	   &  I     &  1.75        &  63.10    &  1.01    &  1.36    &  0.71    &  2.39    &  1.68 \\
V380 Ori 		   &  B9       &  460   &  1.7 	   &  I     &  0.97        &  71.00    &  0.97    &  1.25    &  0.61    &  1.72    &  0.95 \\
BF Ori\dag 		   &  F2       &  460   &  0.0 	   &  I     &  $\leq$0.02  &  3.70     &  1.04    &  1.43    &  0.80    &  1.54    &  1.22 \\
HD 37490\dag 		   &  B2       &  360   &  0.4 	   &  III   &  $\leq$0.01  &  12.52    &  0.01    &  0.12    &  0.04    &  1.44    &  1.01 \\
MWC 137\dag 		   &  B0       &  1300  &  4.5 	   &  I     &  $\leq$0.02  &  14.80    &  0.76    &  1.12    &  0.98    &  1.51    &  2.49 \\
LkH$\alpha$ 215\dag 	   &  B8$^*$   &  800   &  2.1 	   &  I     &  $\leq$0.02  &  26.70    &  0.63    &  0.98    &  0.54    &  1.10    &  1.85 \\
HD 259431 		   &  B2       &  800   &  1.6 	   &  I     &  0.53        &  52.90    &  0.88    &  1.25    &  0.29    &  1.98    &  2.14 \\
R Mon 			   &  B0       &  800   &  4.3 	   &  II    &  3.16        &  93.10    &  1.47    &  2.16    &  0.70    &  2.44    &  2.50 \\
LkH$\alpha$ 25 		   &  A0       &  800   &  0.6 	   &  II    &  1.63        &  58.20    &  1.27    &  1.73    &  1.06    &  3.41    &  3.72 \\
HD 52721\dag 		   &  B1       &  1150  &  1.0 	   &  III   &  $\leq$0.01  &  8.75     &  0.08    &  0.16    &  0.28    &  1.06    &  3.02 \\
LkH$\alpha$ 218\dag 	   &  B9       &  1150  &  1.5 	   &  I     &  $\leq$0.06  &  31.90    &  0.50    &  --      &  --      &  --      &   --  \\
HD 53367\dag 		   &  B0       &  1150  &  2.3 	   &  III   &  $\leq$0.01  &  5.00     &  0.13    &  0.22    &  0.15    &  0.36    &  3.97 \\
NX Pup 			   &  A7$^*$   &  450   &  1.5 	   &  I     &  0.90        &  44.20    &  1.15    &  1.52    &  0.48    &  0.06    &  1.79 \\
HD 76534\dag 		   &  B3       &  870   &  1.0 	   &  III   &  $\leq$0.02  &  14.30    &  0.01    &  0.28    &  0.29    &  0.04    &  2.49 \\
HD 97048 		   &  B9       &  215   &  1.3 	   &  I     &  0.16        &  51.10    &  0.62    &  1.37    &  0.02    &  2.31    &  2.90 \\
HD 150193\dag 		   &  A2       &  160   &  1.5 	   &  I     &  $\leq$0.04  &  7.82     &  0.71    &  1.25    &  0.36    &  2.68    &  1.37 \\
KK Oph 			   &  A6       &  160   &  1.6 	   &  I     &  1.69        &  44.90    &  1.23    &  1.58    &  1.39    &  1.08    &  1.65 \\
HD 163296\dag 		   &  A0       &  160   &  0.3 	   &  I     &  $\leq$0.03  &  22.50    &  0.83    &  1.20    &  0.37    &  2.43    &  1.73 \\
MWC 297 		   &  O9       &  450   &  8.3 	   &  I     &  8.20        &  102.60   &  0.92    &  1.43    &  0.30    &  1.37    &  1.93 \\
VV Ser 			   &  B9       &  440   &  3.0 	   &  I     &  0.44        &  31.80    &  0.90    &  0.95    &  0.04    &  1.90    &  1.31 \\
BD+40$^{\circ}$ 4124 	   &  B2       &  1000  &  3.0 	   &  I     &  1.10        &  31.80    &  0.92    &  1.21    &  0.97    &  1.73    &   --  \\
BD+41$^{\circ}$ 3731\dag   &  B3       &  1000  &  0.9	   &  III   &  $\leq$0.02  &  ABS      &  0.16    &  0.26    &  --      &  --      &   --  \\
HD 200775 		   &  B2       &  600   &  2.0 	   &  I     &  0.14        &  17.70    &  0.66    &  1.20    &  0.40    &  1.19    &  2.67 \\
LkH$\alpha$ 234 	   &  B7$^*$   &  1000  &  3.4 	   &  I     &  0.03        &  52.10    &  1.03    &  1.36    &  0.44    &  2.09    &  3.04 \\
BD+46 3471 		   &  A0       &  900   &  1.0 	   &  I     &  0.45        &  19.70    &  0.89    &  1.20    &  0.48    &  1.26    &  1.56 \\
LkH$\alpha$ 233 	   &  A5       &  880   &  2.6 	   &  II    &  1.02        &  26.40    &  1.52    &  2.06    &  0.48    &  2.77    &  3.18 \\
MWC 1080 		   &  B0       &  1000  &  5.3 	   &  I     &  0.24        &  94.20    &  0.93    &  1.54    &  0.24    &  1.37    &  1.78 \\
PV Cep 			   &  F2       &  500   &  7.0 	   &  II    & 13.80        &  87.20    &  1.15    &  1.78    &  0.52    &  2.8     &   --  \\
Z CMa 			   &  F5       &  1150  &  2.8 	   &  II    &  2.54        &  24.80    &  0.80    &  1.66    &  0.96    &  1.69    &   --  \\
V645 Cyg 		   &  A0       &  3500  &  3.1 	   &  II    &  6.00        &  57.40    &  1.98    &  1.95    &  1.25    &  1.88    &   --  \\
BD+65 1637\dag 		   &  B9       &  1000  &  1.9 	   &  III   &  $\leq$0.04  &  12.71    &  0.03    &  0.51    &  0.41    &  --      &   --  \\
LkH$\alpha$ 134 	   &  B2$^*$   &  700   &  2.1 	   &  --    &  0.50        &  31.54    &  0.85    &  0.70    &  2.35    &  2.57    &   --  \\
\hline 
\end {tabular}
\vspace {1mm}
\caption{The optical and IR data for the Herbig Ae/Be star sample. 
W([OI]) and W(H$\alpha$) are the absolute values of the equivalent
widths of the [OI]$\lambda$6300 and H$\alpha$ lines respectively. The
reddening corrected IR colours are calculated from photometry
published by Hillenbrand et al.\, (1992) and Hamann \& Persson
(1992). The classification by group (Grp.) is also from Hillenbrand et
al.\ (1992). Those stars marked with (\dag) have only {\em upper limits} for
W([OI]). The spectral type of those stars marked with an asterisk (*)
is our own estimate, based on identification of absorption lines,
where this differs from the value in the literature, see Corcoran \& 
Ray (1996). All other spectral types, distance estimates and visual
extinctions are from Hillenbrand et al.\ (1992) and Hamann \& Persson
(1992). H$\alpha$ equivalent widths are with respect to the continuum
level and {\em not} the expected photospheric absorption level.}
\end{center}
\normalsize
\end{table*}

\begin {figure*}[t]
\begin {center}
\leavevmode
\epsfxsize=180mm
\epsfbox{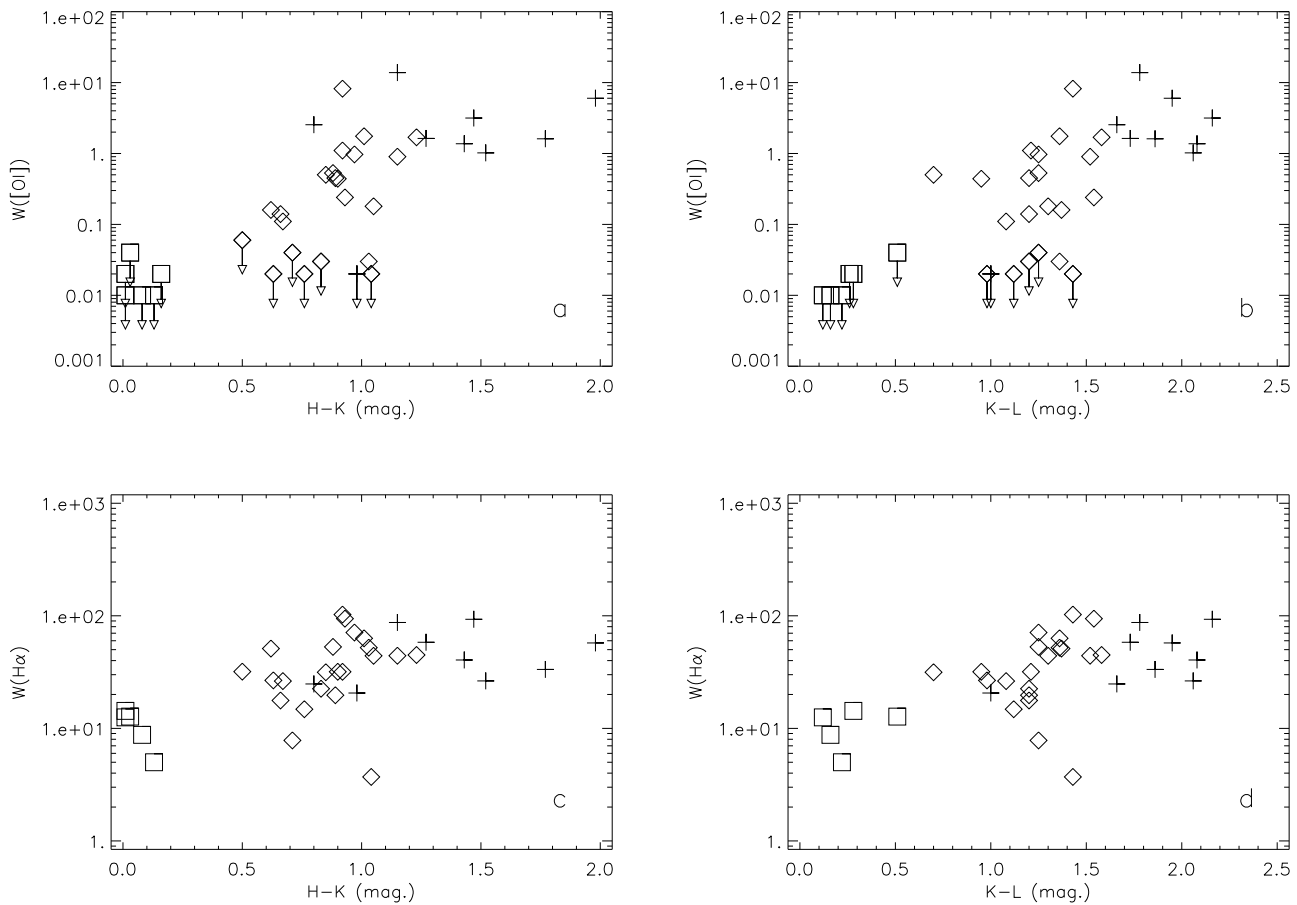}
\caption{Absolute equivalent widths (in angstr\"oms) of [OI]$\lambda$6300 and 
H$\alpha$ plotted against near-infrared colours, H-K and K-L. In this and
subsequent figures, Hillenbrand et al.\ (1992) Group I stars are 
indicated by diamonds, Group II stars by crosses and Group III stars 
by squares. Note that the Group
III stars cluster around the expected IR colours of main sequence
stars, while the Group I and II stars all possess H-K and K-L colours
equal or greater than those expected from an opaque geometrically thin
circumstellar disk. The arrows indicate that the value of \LOI\ is an 
upper limit only, and \OI\ was not detected from that star. A K-L colour
is not known for the Group I star LkH$\alpha$ 218 and so it is missing 
from the plots to the right. Note also that one Group III star, BD+41~3731 
was seen in absorption in H$\alpha$ and is therefore not shown in the 
bottom plots. It is almost certainly not a pre-main sequence star (see text)}
\end {center}
\end {figure*}

\begin{table*}
\footnotesize
\begin{center}
\begin{tabular}{||l||c|c|c|c|c|r|r|r||} \hline
Name                &  V        &  R        &  BC       &  \LOI          &  \LHA      &  \LIRE     &  \LPHOT  &  \LSTAR \\
                    &  (mag)    &  (mag)    &  (mag)    &  (L$_{\odot}$)   &  (L$_{\odot}$) &  (L$_{\odot}$) &  (L$_{\odot}$) &  (L$_{\odot}$) \\
\cline {1-9}
LkH$\alpha$ 198     &  14.32    &  12.93    &  -0.15    & 9.2e-04        & 2.7e-02      &   76.9  &   0.7     &   6.40  \\
V376 Cas            &  15.50    &  14.28    &  -0.08    & 4.1e-04        & 8.4e-03      &   74.6  &   0.5     &   2.90  \\
BD+61 154           &  10.38    &  9.92     &  -0.85    & 1.8e-03        & 4.3e-01      &   131   &   22.4    &   371   \\
Elias 1\dag         &  15.30    &  14.10    &  -0.14    & $\leq$5.0e-06  & 9.5e-04      &   8.40  &   0.1     &   0.80  \\
AB Aur              &  7.07     &  6.82     &  -0.40    & 3.7e-04        & 8.9e-02      &   31.3  &   5.4     &   65.5  \\
HK Ori              &  11.66    &  11.20    &  -0.15    & 1.5e-03        & 5.2e-02      &   15.3  &   1.5     &   13.1  \\
V380 Ori            &  10.37    &  9.95     &  -0.66    & 3.6e-03        & 2.6e-01      &   49.7  &   7.0     &   109   \\
BF Ori\dag          &  11.92    &  11.57    &  -0.06    & $\leq$5.4e-06  & 1.0e-03      &   6.30  &   0.5     &   3.10  \\
HD 37490\dag        &  4.57     &  4.57     &  -2.23    & $\leq$1.5e-03  & 1.7e-00      &   551   &   349.4   &   17900 \\
MWC 137\dag         &  11.84    &  10.74    &  -3.17    & $\leq$1.8e-03  & 1.3e-00      &   639   &   173.1   &   29900 \\
LkH$\alpha$ 215\dag &  10.36    &  9.94     &  -0.85    & $\leq$2.9e-04  & 3.9e-01      &   108   &   41.6    &   573   \\
HD 259431           &  8.70     &  8.40     &  -2.23    & 2.3e-02        & 2.3e-00      &   310   &   110.2   &   5950  \\
R Mon               &  12.80    &  12.17    &  -3.17    & 2.0e-02        & 7.3e-01      &   305   &   21.2    &   3890  \\
LkH$\alpha$ 25      &  12.75    &  12.55    &  -0.40    & 8.0e-04        & 2.9e-02      &   9.20  &   0.9     &   10.5  \\
HD 52721\dag        &  6.62     &  6.63     &  -2.50    & $\leq$2.4e-03  & 2.9e-00      &   1420  &   932.8   &   61600 \\
LkH$\alpha$ 218\dag &  11.87    &  11.55    &  -0.66    & $\leq$2.8e-04  & 1.5e-01      &   36.5  &   9.1     &   142   \\
HD 53367\dag        &  7.02     &  6.65     &  -3.17    & $\leq$6.3e-03  & 3.5e-00      &   3380  &   1456.0  &   261000 \\
NX Pup              &  10.61    &  10.20    &  -0.12    & 2.2e-03        & 1.1e-01      &   44.0  &   4.2     &   42.3  \\
HD 76534\dag        &  7.96     &  7.83     &  -1.77    & $\leq$1.2e-03  & 8.3e-01      &   260   &   122.4   &   5240  \\
HD 97048            &  8.46     &  8.20     &  -0.66    & 4.9e-04        & 1.6e-01      &   21.0  &   5.5     &   95.7  \\
HD 150193\dag       &  8.80     &  8.41     &  -0.25    & $\leq$6.4e-05  & 1.3e-02      &   14.7  &   2.8     &   31.9  \\
KK Oph              &  10.57    &  10.09    &  -0.14    & 6.1e-04        & 1.6e-02      &   8.80  &   0.8     &   6.20  \\
HD 163296\dag       &  6.83     &  6.77     &  -0.40    & $\leq$9.9e-05  & 7.4e-02      &   26.0  &   5.8     &   74.5  \\
MWC 297             &  12.17    &  9.45     &  -3.34    & 3.4e+00        & 4.2e+01      &   2830  &   449.1   &   102000 \\
VV Ser              &  11.87    &  11.21    &  -0.66    & 1.1e-03        & 7.8e-02      &   58.3  &   5.1     &   83.0  \\
BD+40 4124          &  10.54    &  --       &  -2.23    & 5.8e-02        & 1.7e-00      &   1260  &   124.4   &   6190  \\
BD+41 3731\dag      &  9.87     &  --       &  -1.77    & $\leq$2.8e-04  & ABS          &   51.3  &   28.2    &   1090  \\
HD 200775           &  7.36     &  6.88     &  -2.23    & 1.8e-02        & 2.3e-00      &   738   &   308.0   &   16600 \\
LkH$\alpha$ 234     &  11.90    &  10.90    &  -1.04    & 6.5e-03        & 1.1e-00      &   183   &   42.1    &   855   \\
BD+46 3471          &  8.21     &  7.89     &  -0.40    & 2.7e-02        & 1.2e-00      &   299   &   93.7    &   1260  \\
LkH$\alpha$ 233     &  13.56    &  12.62    &  -0.15    & 2.1e-03        & 5.4e-02      &   35.0  &   3.5     &   30.3  \\
MWC 1080            &  11.68    &  10.55    &  -3.17    & 2.5e-02        & 9.8e+00      &   1430  &   235.6   &   42900 \\
PV Cep              &  17.76    &  16.44    &  -0.06    & 4.8e-03        & 3.0e-02      &   41.8  &   1.1     &   10.8  \\
Z CMa               &  9.33     &  8.12     &  -0.04    & 6.4e-01        & 6.3e-00      &   4250  &   335.8   &   2760  \\
V645 Cyg            &  15.00    &  14.10    &  -0.40    & 6.9e-02        & 6.6e-01      &   10600 &   17.6    &   254   \\
BD+65 1637\dag      &  10.08    &  9.53     &  -0.66    & $\leq$1.2e-03  & 3.7e-01      &   131   &   58.9    &   809   \\
LkH$\alpha$ 134     &  12.00    &  --       &  -2.23    & 1.5e-03        & 9.2e-02      &   16.8  &   3.3     &   345   \\
\hline 
\end {tabular}
\vspace {1mm}
\caption {Table of observed broadband V and R magnitudes plus bolometric 
correction (B.C.) applied in calculating the photospheric luminosity, \LSTAR , 
of individual stars (see text). The table also gives the derived line 
luminosities, \LIRE\ the infrared excess (over the range 0.7$\mu$m to 
10.2$\mu$m, see text for precise definitions), \LPHOT\ the expected infrared 
contribution from the stellar photosphere (over the same range) as well as the
photospheric luminosity, \LSTAR , for the Herbig Ae/Be star sample. 
All luminosities are corrected for reddening.
Those stars marked with (\dag) have only {\em upper limits} for
\LOI. ABS indicates the line is in absorption. The bolometric
corrections are taken from Zombeck (1990) and the V and R magnitudes
are taken from Hillenbrand et al.\ (1992) and Hamann \& Persson (1992).}
\end{center}
\normalsize
\end{table*}

The [OI] and H$\alpha$ line luminosities were calculated from the
absolute line equivalent widths and the continuum luminosity in the R
band, centred on 7000\AA, using $4\pi d^2 {\rm F}_{\lambda}{\rm
W}_{{\rm line}}$, where $d$ is the distance to the source,
F$_{\lambda}$ is the flux density of the R band corrected for
extinction and W$_{{\rm line}}$ is the absolute emission line
equivalent width in angstr\"oms.  This method was used as a number of
the observations lacked suitable flux calibration standards that would
allow a direct measurement of the line luminosities. In the case of
three stars (BD+40$^{\circ}$ 4124, BD+41$^{\circ}$ 3731 and
LkH$\alpha$ 134) no R band observations were available and the V
magnitude was used, extrapolated to the R band, e.g.\ from the SEDs in
Hillenbrand et al.\ (1992). Estimated errors for the luminosities are
typically no better than $\pm$20\% depending on the uncertainties in
the visual extinction and actual R band magnitude of the source at the
time of observation. Uncertainties in the distance and visual
extinction estimates can introduce sizable errors, see
Discussion. Estimates of the visual extinction, particularly, vary
greatly in the literature (e.g.\ LkH$\alpha$ 198 with a quoted
A$_{{\rm v}}$ of 2.5 in Hillenbrand et al.\ 1992 and an A$_{{\rm v}}$
of 5 in Berrilli et al.\ 1992) and reflect the uncertainty in the dust
composition and distribution about these stars.

\begin {figure*}[t]
\begin {center}
\leavevmode
\epsfxsize=180mm
\epsfbox{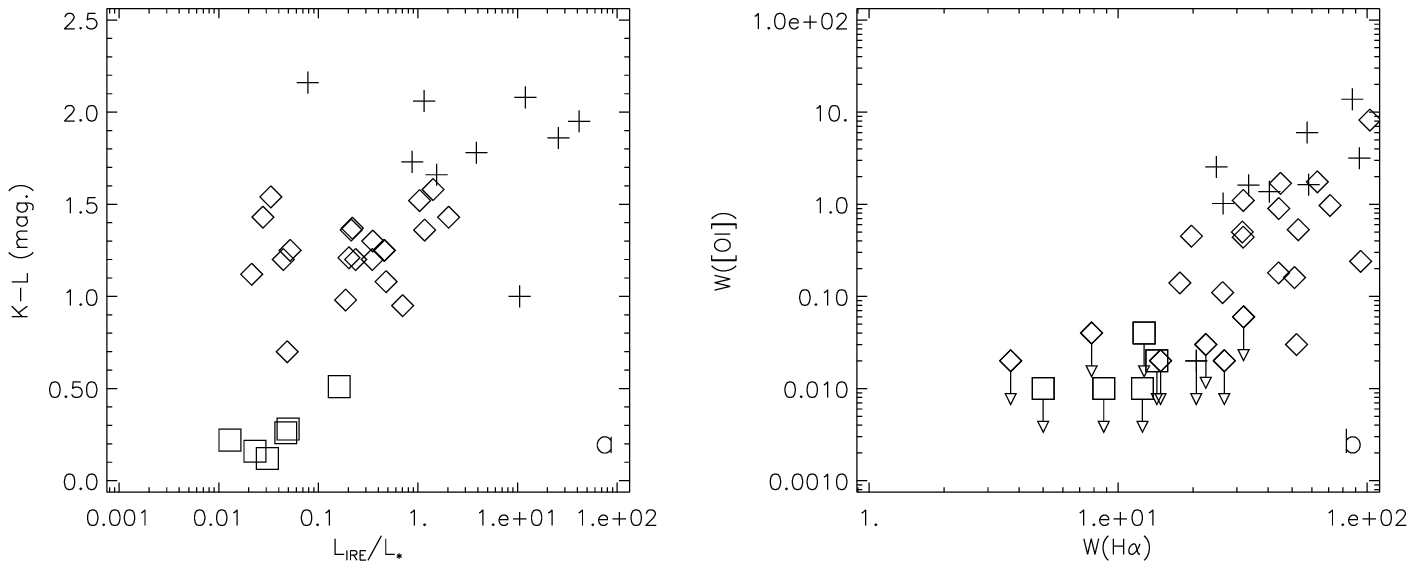}
\caption{a) Near infrared colour K-L plotted against the normalized
infrared excess luminosity, \LIRE/\LSTAR. The infrared colour
increases with increasing near-infrared excess luminosity, as expected
from the connection found by Hartigan et al.\ (1995) between K-L
colour and the veiling luminosity in CTTS. The Group I stars
to the left of the main distribution are discussed in the text. b)
Absolute equivalent widths of \OI\ and H$\alpha$. The forbidden line scales
well with the Balmer line. Those Group I stars without detectable \OI\
(identified in Table 2) also show the weakest H$\alpha$ equivalent
widths of all Group I stars, suggesting reduced activity in the wind
is measured by both the H$\alpha$ and \OI\ lines. The Group II stars
appear to have the highest levels of \OI\ for a given H$\alpha$ which
is in keeping with their less evolved state, more active circumstellar
disks and consequently more powerful outflows.}
\end {center}
\end {figure*}

	Table 2 lists the 37 HAEBES selected from our dataset according to the 
criteria described above. Of these 24 show [OI]$\lambda$6300 emission. The 
remaining 13 stars have no [OI]$\lambda$6300 emission and are marked with a 
(\dag) and the detection limit quoted is an upper limit for W([OI]). 
These will be discussed later. The broad band photometry 
values in Table 2 are taken from Hillenbrand et al.\ (1992) with the exception 
of the stars PV Cep, Z~CMa, V645~Cyg and LkH$\alpha$~134 which are from 
Hamann \& Persson (1992).  All are corrected for interstellar extinction, using
the visual extinctions quoted in Hillenbrand et al.\ (1992) and Hamann
\& Persson (1992), assuming the standard interstellar reddening curve 
with $A_V/E_{B-V} \approx 3.1$, and that the ratio of the absorption 
values, $A_{\lambda}/A_V$ scale, with wavelength, as tabulated in Scheffler \& 
Els\"asser (1982). Table~3 presents the calculated luminosities of \OI\ and
H$\alpha$, along with the infrared and stellar (photospheric) luminosities 
for the 37 HAEBES. The V and R magnitudes of the stars as well as the 
bolometric correction applied in each case is also listed. 

\begin {figure*}[t]
\begin {center}
\leavevmode
\epsfxsize=180mm
\epsfbox{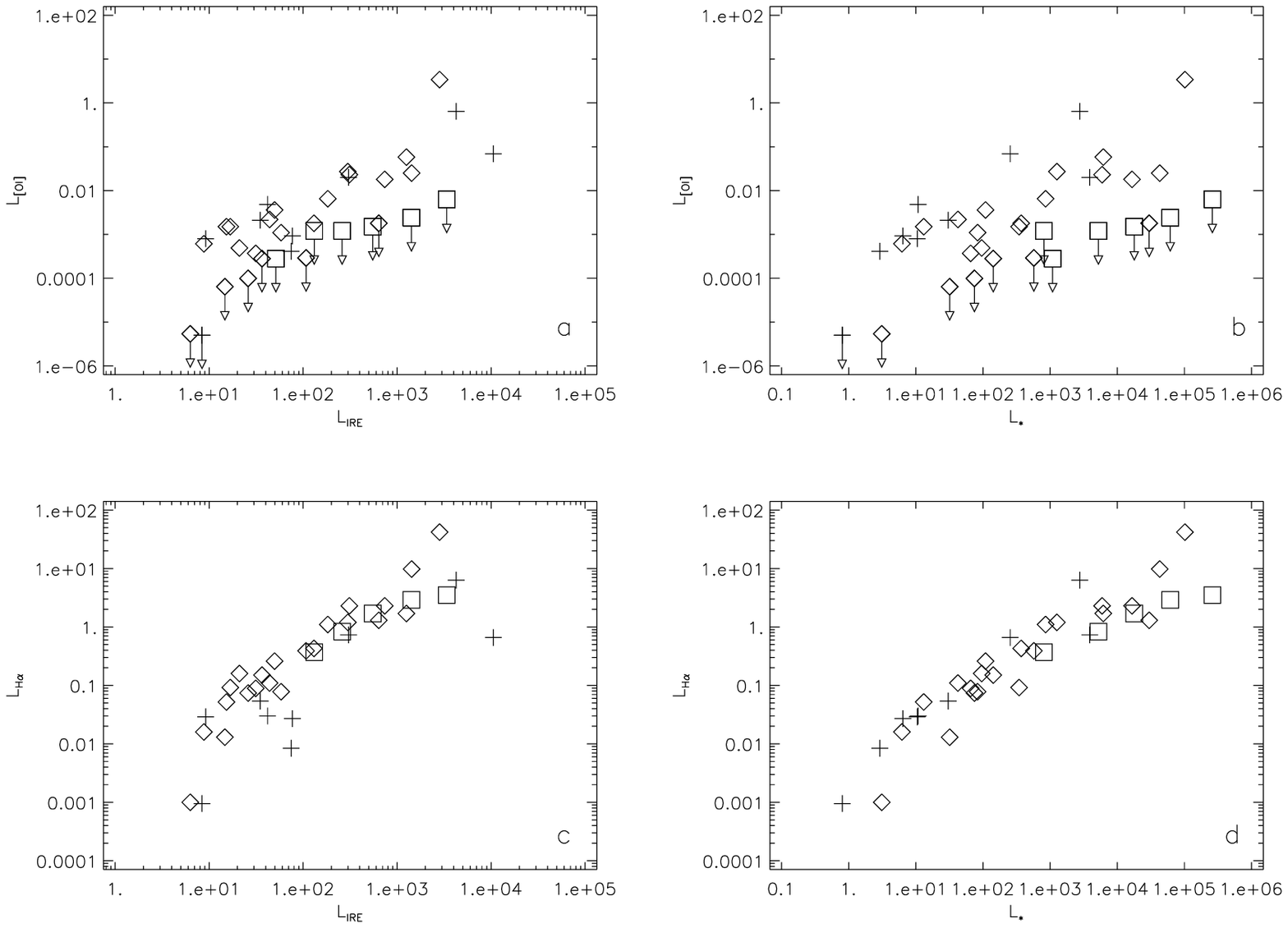}
\caption{a) Emission line luminosity of [OI]$\lambda$6300 plotted
against the infrared excess, \LIRE. b) Emission line luminosity of [OI]
$\lambda$6300 plotted against photospheric luminosity, \LSTAR. c)
Emission line luminosity of H$\alpha$ plotted against the infrared
excess (\LIRE). d) Emission line luminosity of H$\alpha$ plotted
against the photospheric luminosity, \LSTAR. All luminosities are in
units of L$_{\odot}$. See Fig.\ 1 for key.}
\end {center}
\end {figure*}

\begin {figure*}[t]
\begin {center}
\leavevmode
\epsfxsize=180mm
\epsfbox{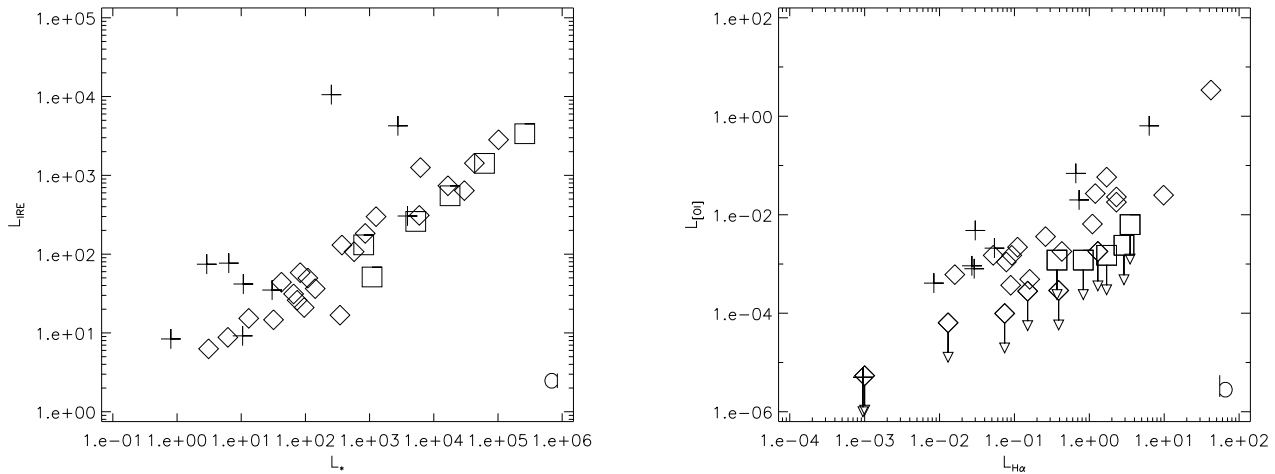}
\caption[garbage caption]{a) The infrared excess luminosity, \LIRE\
vs.\ the photospheric luminosity, \LSTAR. b) Emission line luminosity
of [OI]$\lambda$6300 vs.\ H$\alpha$. All luminosities are in units of
L$_{\odot}$. See Fig.\ 1 for key.}
\end {center}
\end {figure*}

\begin {figure*}[t]
\begin {center}
\leavevmode
\epsfxsize=110mm
\epsfbox{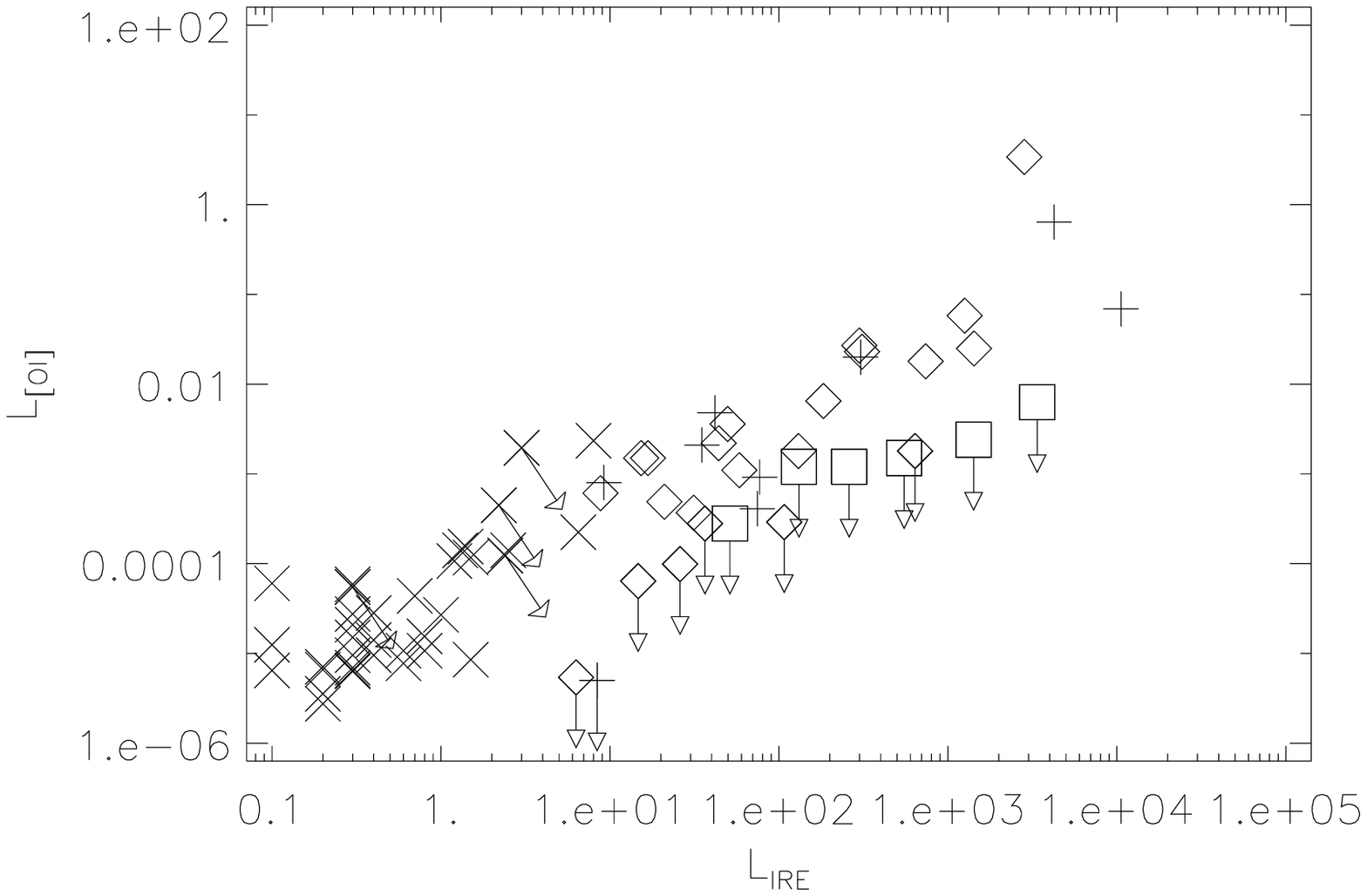}
\caption[Comparison with the T-Tauri Stars] {Emission line luminosity
of [OI]$\lambda$6300 plotted against the infrared excess (\LIRE) for
our sample of Herbig Ae/Be stars and the T-Tauri stars from Cabrit et
al.\ (1990). All luminosities are in units of L$_{\odot}$. TTS are
represented by x's, all other stars are represented as described in
Fig.\ 1. Those TTS with arrows are cont/e sources as described in
Cabrit et al.\ (1990) and may have incorrectly calculated visual
extinctions.}
\end {center}
\end {figure*}

\section{Results}

\subsection{Relative wind strength and near-IR colours}

	A number of infrared colours were chosen to sample the dust
temperature gradient and, indirectly, the accretion activity; H-K,
K-L, L-M, M-N, and IRAS 12 and 25$\mu$m fluxes converted to magnitudes
(see Table 2). The near-infrared photometry (H; 1.65$\mu$m, K; 2.2$\mu$m,
L; 3.6$\mu$m, M; 4.8$\mu$m, N; 10.2$\mu$m) and the IRAS observations
at 12$\mu$m and 25$\mu$m effectively sample different temperature
regimes of the circumstellar dust, ranging from about 1700K (i.e.\
close to the dust sublimation point) at 1.65$\mu$m to about 100K at
25$\mu$m (these are approximate values from Wein's displacement
law). The IR colours are accurate to approximately $\pm$ 0.02 mag.

	Fig.\ 1 plots the relationship between the absolute equivalent widths
of [OI]$\lambda$6300 and H$\alpha$ and the near-IR colours H-K and
K-L, given in Table 2.  The HAEBES are clearly separated
into three distinct regions of the plot depending on their Hillenbrand et
al.\ (1992) group (see Table 2). Here we recall that Hillenbrand et al.\ 
(1992) classified HAEBES by the slope of their near-infrared spectral energy 
distribution (SED). Group I stars
show SEDs with a declining slope in the near-infrared that can be
modeled with a geometrically flat optically thick disk with a T
$\propto r^{-3/4}$ temperature law. The Group II stars have SEDs with
flat or rising slopes in the near to mid-infrared, and the Group III
Herbig stars have near-infrared SEDs with little or no deviation from
a normal stellar spectrum. The Group III stars in our sample all show
values of H-K and K-L close to the values expected for main-sequence
B-type stars ($\sim 0.0$) and no detectable [OI]$\lambda$6300
emission. Their distinct separation from the Group I \& II stars in Fig.\
1 is analogous to the division between the weak-line T Tauri stars (WTTS) 
and the CTTS in the corresponding plots for lower mass stars (Edwards et 
al.\ 1993). All of the Group III stars in our sample 
show absolute equivalent widths of H$\alpha \leq$ 15\AA. The possibility of the
Group III stars being an intermediate mass analogue to the WTTS 
is discussed later. The cut-off colour indices for the Group I stars are H-K 
= 0.4 and K-L = 0.8 in line with the Hillenbrand et al.\ (1992) definition 
of this group (see also \S 4.1).  The transition 
from Group I to Group II stars is then one of increasing redness as the 
relative contribution from the near-infrared to the SED goes up. 

A clear result in Figs.\ 1a and 1b is that the equivalent width of [OI] 
emission, W([OI]), scales with H-K and K-L and, in particular,  the 3 groups 
have very different [OI] emission line properties. While the Group III stars, 
i.e.\ those without optically thick disks (H-K $<$ 0.4 and K-L $<$ 0.8), 
have no detectable [OI] emission, the equivalent width of the [OI] emission 
for Group I stars rises by over 2 orders of magnitude with increasing K-L. The 
Group II YSOs then constitute those stars with the largest [OI] equivalents 
widths. The results for the W(H$\alpha$) vs.\
H-K \& vs.\ K-L (Figs.\ 1c, d) are found to be comparable to those for
W([OI]). A similar effect was discovered by Ghandour et al.\
(1994) for a smaller sample of HAEBES when looking for a correlation between 
the equivalent width of \OI\ and I-N colour. As Ghandour et al.\ (1994) were 
comparing the ratio of the flux at 0.9$\mu$m (I) and 10.6$\mu$m (N) they did not
find as clear a separation (for reasons that will be discussed shortly) 
between the Group III stars and the Group I
and II stars as we do here. 

A very interesting finding, if we compare Figs.\ 1b and 1d with Fig.\ 2 of 
Edwards et al.\ (1993), is that the dependence of both [OI] and H$\alpha$ 
equivalent width on K-L is {\em virtually identical} in both the HAEBES and
CTTS. As in the CTTS, the absolute value of the [OI] equivalent width is found 
to vary in the HAEBES by approximately two orders of magnitude, from 0.1 to 
10\AA, as K-L is increased from 0.5 to 1.5 while the corresponding relative 
change in the H$\alpha$ equivalent width, from 10 to about 100\AA, is smaller, 
only one order of magnitude. Such close parallels in phenomenological 
relationships argue strongly that the same basic accretion/outflow process is 
operating in both groups of stars. 

The good degree of correlation between [OI]$\lambda$6300 equivalent
width and near-IR colours for the HAEBES does not extend to colours longward of
$\sim$4.0$\mu$m.  Moreover in the corresponding plots at longer wavelengths, 
the 3 Hillenbrand et al.\ (1992) groups do not occupy  distinct regions {\em 
of IR colours} as in Fig.\ 1. For example, comparisons of the W([OI]) 
vs.\ L-M show little correlation of the two variables. 
The circumstellar conditions giving rise to the L-M, M-N
and IRAS 12$\mu$m~-~25$\mu$m colours must depart significantly from
the conditions in the inner region (as indicated by the H-K and K-L
colours).  Note that the material emitting at $\sim 4\mu$m is
typically situated at about an AU from an A0 star (assuming a flat
optically thick reprocessing disk with an $r^{-3/4}$ temperature law). 
[Table 4, which will be discussed later, shows the results of two
survival analysis tests on the W([OI]) vs.\ IR colours
data. Both tests agree with the visual correlations seen
in Fig.\ 1 and confirm the lack of correlation between W([OI]) and
L-M, M-N \& IRAS 12-25$\mu$m colours.]

\begin {table*}
\footnotesize
\begin{center}
\begin{tabular}{||l|c|c|c|c||} \hline
Plotted Variables & \multicolumn{2}{c|}{Gen. Kendall's Tau} &
  \multicolumn{2}{c||}{Cox's Prop. Hazard} \\ 
& Z-value & Prob.&  $\chi^2$ & Prob. \\
\hline
W([OI]) vs. H-K         & 4.47    & $<$0.0001    & 23.90 & $<$0.0001 \\
W([OI]) vs. K-L         & 4.74    & $<$0.0001    & 22.13 & $<$0.0001 \\
W([OI]) vs. L-M         & 2.09    & 0.0371       & 3.43  & 0.0641    \\
W([OI]) vs. M-N         & 1.60    & 0.1107       & 3.02  & 0.0824    \\
W([OI]) vs. 12$\mu$m - 25$\mu$m & 0.019 & 0.9847 & 0.002 & 0.9678    \\
W($H_{\alpha}$) vs. H-K & 3.73    & 0.0002       & 15.94 & 0.0001    \\
W($H_{\alpha}$) vs. K-L & 3.88    & 0.0001       & 15.13 & 0.0001    \\
W($H_{\alpha}$) vs. L-M & 0.86    & 0.3900       & 0.93  & 0.3338    \\
W($H_{\alpha}$) vs. M-N & 1.41    & 0.1588       & 2.73  & 0.0984    \\
W($H_{\alpha}$) vs. 12$\mu$m - 25$\mu$m & 0.078 & 0.94   & 0.005 & 0.9423 \\
K-L vs.\ \LIRE/\LSTAR\  & 3.38    & 0.0007       & 11.89 & 0.0006    \\
\hline
\LOI\ vs.\ \LIRE\       & 3.67    & 0.0002       & 22.19 &$<$0.0001  \\
\LHA\ vs.\ \LIRE\       & 4.96    &$<$0.0001     & 6.89  & 0.0087    \\
\LOI\ vs.\ \LSTAR\      & 2.31    & 0.209        & 0.03  & 0.8575    \\
\LHA\ vs.\ \LSTAR\      & 5.53    &$<$0.0001     & 2.81  & 0.0936    \\
\LIRE\ vs.\ \LSTAR\     & 3.87    & 0.0001       & 3.39  & 0.0655    \\
\LOI\ vs.\ \LHA\        & 3.13    & 0.0017       & 1.37  & 0.2415    \\
\hline
\LOI\ vs.\ M$_{{\rm cs}}$   & 3.02  & 0.0025     & 8.59  & 0.0034    \\
\LIRE\ vs.\ M$_{{\rm cs}}$  & 2.97  & 0.0030     & 19.14 & $<$0.0001 \\
\LSTAR\ vs.\ M$_{{\rm cs}}$ & 1.85  & 0.0645     & 0.87  & 0.3503    \\
\hline
\end {tabular}
\vspace{1mm}
\caption [Statistical Tests on the Correlations Sample] {The results
of the statistical survival tests. All probabilities are {\protect\it
those of no correlation existing} between the compared variables. For
comparison the probabilities of no correlation existing for the sample
of 36 TTS in the sample of Cabrit et al.\ (1990) is $\leq$ 0.0001 for
the comparison of \LOI\ and \LHA\ against both \LIRE\ and \LSTAR. The
luminosity range sampled by Cabrit et al.\ (1990) was of course smaller.}
\end{center}
\normalsize
\end{table*}

	Fig.\ 2a plots K-L colour against the normalized infrared excess,
\LIRE/\LSTAR. \LIRE\ and \LSTAR\ are listed in Table 3 and are defined
more precisely below, but can be taken as the excess IR luminosity roughly
over the range 1--10$\mu$m and the stellar bolometric luminosity, as determined 
using the reddening corrected visual magnitude, respectively. From Fig.\ 2a 
it can be seen that K-L colour is correlated with \LIRE/\LSTAR\ and 
that the correlation extends to values of \LIRE/\LSTAR\ well beyond the 
expected regime for pure reprocessing disks. Moreover an identical
correlation (e.g., Edwards et al.\ 1993) with approximately the same slope 
and a similar range in (K-L, \LIRE/\LSTAR) space is found for the T Tauri stars 
(TTS). This clearly suggests that, as with the TTS, K-L is a measure  
of the normalized accretion luminosity (\LACC/\LSTAR).

In Fig.\ 2b we have plotted our relative measure of wind strength, 
W([OI]), against W(H$\alpha$). Not surprisingly the two equivalents widths 
appear correlated and the correlation is exactly the same as found with the 
CTTS (e.g., Cabrit et al. 1990) with the same typical ratio between
the [OI] and H$\alpha$ line strengths. In particular Fig.\ 2b also shows that 
those Group I stars with {\em no} detected \OI, have low absolute H$\alpha$ 
equivalent widths. 

\subsection{Luminosity measurements}

For the purposes of the analysis here, a number of quantities are
defined. The observed near-IR luminosity, L$_{{\rm nir}}$, is the
luminosity of a given star over the spectral range 0.7$\mu$m -- 10.2$\mu$m. 
The corresponding photospheric luminosity, \LPHOT, is calculated as the 
integrated emission from a blackbody of temperature T$_{eff}$
and radius as given by Hillenbrand et al.\ (1992). Consequently, the
infrared excess, \LIRE, is = L$_{{\rm nir}}$ - \LPHOT. This is the
value we use as a diagnostic of the accretion luminosity. The stellar
luminosity, \LSTAR\ quoted in Table 3 is calculated from the reddening
corrected visual magnitude of each star, with visual extinctions from
Hillenbrand et al.\ (1992) and Hamann \& Persson (1992), using
\LSTAR/${\rm L}_{\odot} = 10^{-0.4({\rm M}_{{\rm bol}}-4.72)}$
and main-sequence bolometric corrections (Zombeck 1990). 

All the L$_{{\rm nir}}$ luminosities were calculated, assuming
spherically isotropic emission, by a trapezoidal integration of the
published photometry (Hillenbrand et al.\ 1992; Hamann \& Persson 1992)
over the spectral range covered by the bands R to N (roughly 0.7$\mu$m
-- 10.2$\mu$m)\footnote[1]{This wavelength range was chosen because
any signature of accretion is more likely to be dominant in the
near-infrared whereas one expects a proportionally increased 
contribution from reprocessed light at longer wavelengths. This should
be particularly notable in the case of a flared disk
(Kenyon \& Hartmann 1987) where the contribution from reprocessed
light at 60$\mu$m with a surface scaling as $R^z$ can be increased by a
factor of 10, for z = 9/8 or a factor of 33, for z = 5/4, over that
from a flat disk (z=0).}.

	The results of the various luminosity correlation
investigations are plotted in Figs.\ 3, 4 and 5. Figs.\ 3a, b compare
the [OI]$\lambda$6300 emission line luminosity with \LIRE\ and \LSTAR\
respectively to determine the relationship between our diagnostic of 
wind mass-loss rate with the accretion and stellar luminosities respectively. 
Figs.\ 3c, d are the corresponding plots for the H$\alpha$ line luminosity. 
Fig.\ 4 shows the \LIRE\ vs.\ \LSTAR\ relationship and the comparison of the two
emission line luminosities against each other. Fig.\ 5 is a combined plot
showing \LOI\ line luminosity against  \LIRE\ for our sample of
HAEBES and the TTS sample of Cabrit et al.\ (1990). 

The data points in Fig.\ 3 are separated by group (according to
Hillenbrand et al.\ 1992), with downward arrows indicating those stars
with upper limits to the observed [OI]$\lambda$6300 emission. One can
see that there is a clear relationship between \LOI\, our wind tracer, \LHA, 
and \LIRE\ (Figs.\ 3a and c). As discussed in the Introduction, we expect,
and find, that \LOI\ and \LHA\ also correlate with \LSTAR\, although 
the correlations (Figs.\ 3b and d) are not as good as with \LIRE 
(see also the results of the statistical tests which are described below). 
Note, once again, that all of the Group III stars show no detected
[OI]$\lambda$6300 emission and have only upper limits plotted in Fig.\
3.  Fig.\ 4a shows the expected correlation between 
\LIRE\ and \LSTAR\ although we note that the Group II stars show the 
greatest scatter in the plot reflecting their wide range of accretion
luminosities for a given \LSTAR. In Fig.\ 4b we see that \LOI\ is 
typically 1\% of \LHA\ underlying the point, already made by Fig.\ 2b,
that, while we might expect some forbidden line emission even from those
stars with weak H$\alpha$ emission, it will be difficult to detect. 
Fig.\ 5, shows a plot of \LOI\ against \LIRE\ again but this time we have 
included the corresponding data for 36 TTS from Cabrit et al.\ 
(1990). Note that the \LIRE\ calculated here is equivalent to the near to 
mid-IR excess luminosity, L$_{{\rm mir}}$ as defined in Cabrit et al.\ (1990).
Comparing the data for the two classes of stars we find a smooth
continuation of the correlation between the two variables over a
luminosity range of 5 orders of magnitude.  The mass range sampled
covers $\sim 0.5{\rm M}_{\odot}$ to approximately 10 M$_{\odot}$. 
In Cabrit et al.\ (1990) and Fig.\ 5 there are four TTS marked with
arrows indicating that they are continuum sources (cont/e) in Cabrit
et al.\ (1990) which may have incorrectly calculated visual
extinctions.  Cabrit et al.\ (1990) note that decreasing their $A_{\rm
v}$ by 1 moves them in the direction indicated by the arrows,
reducing the degree to which they deviate from the trend shown by the
HAEBES and the TTS. 

The visual impression of correlations is supported by two statistical
survival analysis tests. The Generalized Kendall's Tau and Cox's
Proportional Hazard tests determine the probability that a random set
of uncorrelated (x,y) points show the same observed degree of
correlation (see, for example, Isobe et al.\ 1986).  The tests deal
with censored bivariate data and, therefore, the stars with no
detected [OI]$\lambda$6300 emission are included in the set as having
line emission equal to the upper limits determined from the minimum
detection level in each case. The results of the survival analysis
tests are shown in Table 4, and indicate that the likelihood of the
observed correlation occurring by chance for \LOI\ and \LHA\ vs.\
\LIRE\ is very small. 

It is interesting to note that the tests show the \LHA\ vs.\ \LIRE\ correlation 
and the \LHA\ vs.\ \LSTAR\ correlation are statistically equally probable,
whereas the \LOI\ vs.\ \LIRE\ correlation is much tighter than the 
correlation between \LOI\ and \LSTAR . We had already noted this to be
the case on the basis of visual inspection. The fact that the relationship
between the line luminosity \LOI\ and the stellar luminosity is weaker
than that between \LOI\ and \LIRE\, clearly suggests that \massloss\ depends 
on \accrate. In particular, the root of the larger scatter in the \LOI\ vs.\ 
\LSTAR\ correlation lies in the huge increase in W[OI] by over two
orders of magnitude with increasing K-L. This creates a separation between the
3 groups at a given \LSTAR, in the sense:  \LOI\ (Group~II) $>$  \LOI\ (Group~I)
$>$  \LOI\ (Group III). Conversely, the
much smaller difference in tightness between \LHA\ vs.\ \LIRE\ and
\LHA\ vs.\  \LSTAR\ is a direct consequence of the much smaller span in
W(H$\alpha$) between the 3 groups, compared with W[OI].

\subsection{Millimeter measurements}

A total of 16 Herbig Ae/Be stars with [OI] measurements were observed
with the CSO in 1992. Two of the stars, LkH$\alpha$ 257 and
LkH$\alpha$ 259, lacked complete optical and near-infrared photometric
data and are therefore not in Tables 2 and 3.  Henning et al.\ (1994)
made millimeter observations of an additional 2 stars for which we
have [OI] data. Table 5 gives the combined sample of 18 stars, with
the relevant continuum flux at either 1.1 mm or 1.3 mm depending on
the observations. By assuming that any mm emission is optically thin,
it is possible to estimate an upper limit for the mass of the circumstellar
material, $ M_{{\rm cs}}$, (see, for example, Mannings 1994).

\begin {equation}
	M_{{\rm cs}} \approx \frac {F_{\lambda}d^2}{\kappa_{\lambda}B_{\lambda}(T)}.
\end {equation}

\noindent Here $F_{\lambda}$ is the flux density at wavelength $\lambda$, $d$ 
the distance to the source, $B_{\lambda}(T)$ the Planck function and  
$\kappa_{\lambda}$ is the mass absorption coefficient. $\kappa_{\lambda}$ is
taken to be a power law characterized by $\kappa_{\lambda} =
\kappa_0\left(\frac{\lambda_0}{\lambda}\right)^{\beta}$. The value of
$\kappa_0$ is assumed to $\sim$ 0.1 ${\rm cm}^2{\rm g}^{-1}$ at
$\lambda_0 = 0.25$ mm, as used by Mannings (1994), Beckwith et al.\
(1990) and Beckwith \& Sargent (1991).  Following Mannings (1994) we take the 
value of $\beta \sim 1.$\footnote {Note however that Pollack et al. (1994)
suggest that $\beta \sim 1.5$ over the millimeter range. Combined with
their values for $\kappa_0$, the resulting $\kappa_{\lambda}$ would be
a factor of approximately 2 smaller, resulting in somewhat larger estimates 
of the circumstellar masses. The calculated masses may need to be revised
upward should the work of Pollack et al.\ (1994) prove more
accurate.} The masses quoted in Table 5 assume a
gas to dust ratio of 100:1 (this factor is contained in the normalization 
of $\kappa_{\lambda}$) and are thus total gas+dust masses. 
A dust temperature of 37~K was assumed in all cases (Mannings 1994).
These mass estimates are subject to some variation due to the
differing beam sizes used in the different observations.  Henning et
al.\ (1994) used a 23$''$ beam, as opposed to the 30$''$ CSO beam. The
difference in beam size will result in varying amounts of background
contamination in the flux measurements and consequently overestimates
of the target's circumstellar mass. This effect is most pronounced for
the most distant stars where the region sampled is much larger than
that for the nearer stars.  For example, V645 Cyg is calculated as
having a circumstellar mass of nearly 90M$_{\odot}$, which is clearly
an overestimate.  To compensate for this we assumed a 1/r$^2$ density
law for the dust and corrected the 1.1mm emission from V645 Cyg to the
average distance of the rest of the sample.  Note that even with this
compensating $r^{-2}$ factor the circumstellar mass of V645 Cyg is
still anomalously high. At the distance of V645 Cyg (3.5 kpc) the 30$''$
beam encompasses nearly a 0.5 pc diameter. There may be multiple
sources in this region (see Corcoran \& Ray 1997b), each of which might
contribute to the 1.3 mm flux. Table 5 shows the two values for the
V645 Cyg circumstellar mass. The masses listed in Table 5 are best
treated as very rough estimates only although we note that for the
many stars we have in common, our mass estimates agree with those 
of Mannings (1994). There are several reasons why we should treat 
the quoted values with suspicion: firstly as we have already stated 
opacity values at millimeter wavelengths are at least uncertain by a
factor of 10 and besides they may vary from place to place
(Men'shchikov and Henning 1997). Moreover assuming the dust distribution
around a young stellar object (YSO) is isothermal is clearly
artificial, one should use much more sophisticated models 
(Di Francesco et al.\ 1997).
\begin{table*}
\footnotesize
\begin{center}
\begin{tabular}{||l||c|c|c|c||} \hline
Star  & SpTy & Distance (pc) & ${\rm F}_{1.1mm}$ (Jy) & M$_{cs}$ ($M_{\odot}$) \\
\cline{1-5}
LkH$\alpha$ 198		& A5 	& 600 	& 0.19 & 0.21   \\
V376 Cas 		& F0 	& 600 	& 0.07 & 0.08   \\
Elias 1 		& A6 	& 160 	& 0.30 & 0.02   \\
AB Aur 			& A0 	& 160 	& 0.13 & 0.01   \\
MWC 137 		& B0 	& 1300 	& 0.08 & 0.41   \\
LkH$\alpha$ 215		& B5 	& 800 	& 0.05 & 0.10   \\
R Mon			& B0 	& 800 	& 0.07 & 0.13   \\
HD 97048$^*$ 		& B9 	& 140 	& 0.45$^*$ & 0.04   \\
KK Oph$^*$		& A6 	& 400 	& 0.52$^*$ & 0.40   \\
HD 163296 		& A0 	& 160 	& 1.21 & 0.09   \\
MWC 297 		& O9 	& 450 	& 0.69 & 0.42   \\
BD+40$^{\circ}$ 4124 	& B2 	& 1000 	& 0.50 & 0.15   \\
MWC 1080 		& B0 	& 1000 	& 0.21 & 0.63   \\
PV Cep 			& A5 	& 500 	& 0.59 & 0.44   \\
Z CMa 			& F5 	& 1150 	& 0.45 & 1.79   \\
V645 Cyg 		& A0 	& 3500 	& 2.25 & 15.8/83.0 \\
LkH$\alpha$ 257 	& B5 	& 900 	& 0.16 & 0.40   \\
LkH$\alpha$ 259 	& A9 	& 850	& 0.13 & 0.02   \\
\hline 
\end {tabular}
\vspace {1mm}
\caption [Millimeter continuum measurements and circumstellar gas+dust
mass estimates]{Millimeter continuum measurements and circumstellar
gas+dust mass estimates.  The data is from CSO observations at 1.1mm,
with the exception of the two stars marked with an asterisk (*) which
are 1.3mm observations from Henning et al.\ (1994). The mass is
calculated from Equation 4 and the error is of order 20\%.  There are
two mass estimates for V645 Cyg (see text for details).}
\label{table2:csodata}
\end{center}
\normalsize
\end{table*}

\begin {figure*}[t]
\begin {center}
\leavevmode
\epsfxsize=180mm
\epsfbox{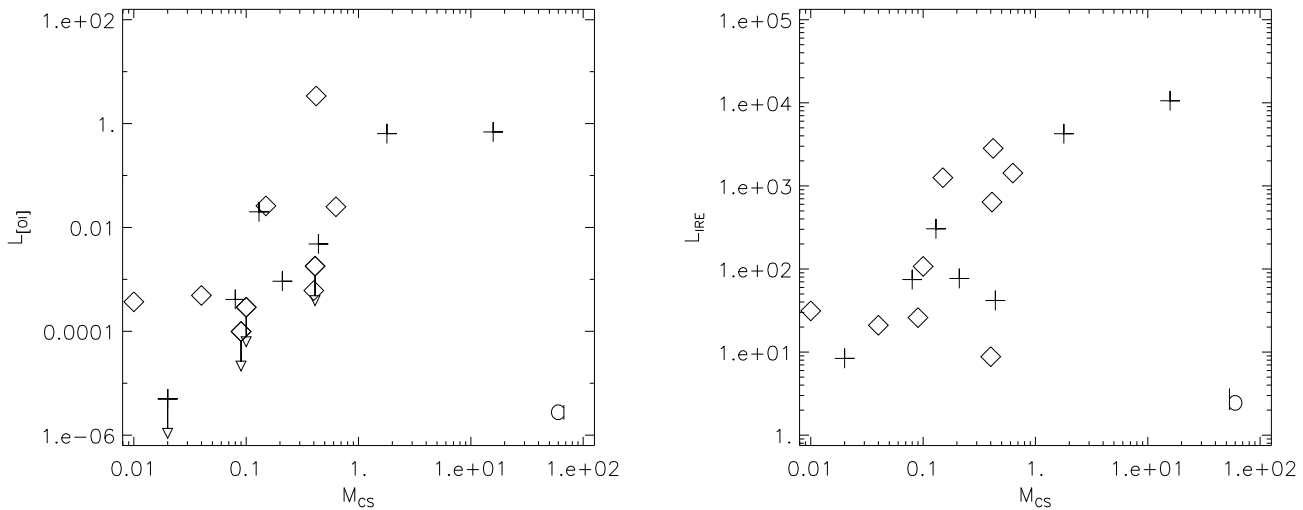}
\caption{Emission line luminosity of [OI]$\lambda$6300 (a) and \LIRE\ (b)
plotted vs.\ the calculated circumstellar mass of gas and dust, in
units of L$_{\odot}$ and M$_{\odot}$ respectively. Symbols are as in
Fig.\ 1.}
\end {center}
\end {figure*}

	Figs.\ 6a and b show plots of the observed [OI] luminosity
and \LIRE\ against derived circumstellar mass. Although the sample is 
relatively small, there appears to be a weak correlation between \LOI\ and 
M$_{{\rm cs}}$ and \LIRE\ and M$_{{\rm cs}}$. We will consider the 
possible origin of this correlation in our Discussion. 

\section{Discussion}

\subsection{Dust colours}

	The three groups of HAEBES, as defined by the shape of the
spectral energy distributions (SEDs) (Hillenbrand et al.\ 1992) fall
into distinct regions in the W([OI]),
W(H$\alpha$) vs.\ H-K, K-L plots. The correlation observed
between the relative strength of the forbidden emission and the near-IR colours
for Groups I and II indicate a possible relationship between the
relative strength of the wind and the temperature gradient in the
circumstellar material, with the equivalent width of \OI\ in Group I
varying by over 2.5 orders of magnitude with increasing K-L. This
cannot be easily understood in terms of purely stellar processes, and
points to some influence of circumstellar environment on the formation
of the line. A plausible explanation for this behaviour, as we shall
now demonstrate, is the presence of an optically thick disk (or a
combination of disk plus halo) around Group I and II stars.

	The expected near-IR colours from an optically thick, passive
reprocessing disk or flat accretion disk (Bertout et al. 1988) can be
calculated simply from the assumption of a $\lambda{\rm F}_{\lambda}
\propto \lambda^{-4/3}$ relationship between the flux density and
wavelength:
\begin {equation}
\frac{f_{\lambda}(m_H)}{f_{\lambda}(m_K)} \approx 
(\frac{\lambda_H}{\lambda_K})^{-7/3}
\end{equation}
\noindent where $f_{\lambda}(m_x)$ is the spectral irradiance
in erg~${\rm cm^{-2}s^{-1}}$\AA$^{-1}$ for an object of magnitude $m_x$
in the $x$ filter. Here $\lambda_H$ = 1.65$\mu$m and $\lambda_K$ = 
2.2$\mu$m are the central wavelengths of the H and K filters respectively. 
Using the zero magnitude irradiance figures provided in Zombeck (1990), one 
finds a value of m$_H$ - m$_K$ = 0.4, or in commonly used notation, H-K = 0.4. 
Similar calculations for K-L yield a value of 0.82. For most of the Group~I 
stars, and all of the Group II stars, the values of H-K, K-L exceed those
possible from a flat reprocessing disk and three obvious explanations are
possible. Firstly the disk could be flared in some manner increasing the
amount of intercepted flux (see, for example, Kenyon \& Hartmann 1987)
and flattening or raising the 1.65--3.6 $\mu$m portion of the SED. 
Flaring of the disk seems to us to be an unlikely explanation because of
geometry: even in a very flared disk, the inner portion (r $\leq$ 10
AU; $\lambda \approx$ 1--10 $\mu$m) should remain relatively flat in
comparison to the outer disk which can intercept significantly more
stellar photons. Another possibility is the presence of an unresolved IR 
companion that distorts the true near-infrared SED of the star.
While this might be important in a small number of cases, recent
searches in the infrared for HAEBES binaries (e.g., Leinert et al.\
1997; Pirzkal et al.\ 1997) show that any companions are usually  
faint in comparison to the HAEBES and so cannot explain their 
infrared excesses. The final possibility is that the increased reddening
is due to a dusty halo which scatters starlight back onto the disk and heats 
it at large radii (Natta 1993; Mannings 1994; Di Francesco et al.\ 1997). 
If one accepts that YSOs with halos
plus disks tend to accrete at higher rates than comparable stars with disks 
alone, then the halo plus disk model explains not only the findings of 
Hartigan et al.\ (1995), who showed a correlation between the infrared colours 
and veiling in TTS but, by extension, our result that the relative
strength of the forbidden line emission is correlated with infrared colour
in HAEBES in the same way. 

Looking at Figs.\ 1a and c, there are, however, a small number of stars
(approximately half a dozen) that do not fit the general trend, i.e.\
they have very red colours but no detected forbidden line emission.
Almost exclusively, they are Group I stars. In some cases,
we may not be dealing with a bona fide HAEBES. For example, HD~163296
(MWC~275) is not in a star forming region, it does not appear to be
associated with any reflection nebulosity, and its spectral
characteristics are inconsistent with it being a HAEBES (Th\'e et al.\
1985). Di Francesco et al.\ (1997) question whether MWC~137, originally
miss-classified as a planetary nebula, is, in fact, a pre-main-sequence star. 
In some cases variability of the forbidden line emission may be the reason 
why it was not detected by us. For example, although we did not 
observe [OI]$\lambda$6300 emission from LkH$\alpha$~218, B\"ohm and Catala 
(1994) did at 3 times our detection level. Variability in forbidden line 
emission strength, by factors of 200--300\% has also been found in a number 
of other HAEBES (B\"ohm and Catala 1994; Corcoran and Ray 1997a).  
Irrespective of their infrared colours, Fig.\ 2b shows that the Group~I
stars with no observed forbidden line emission, are also weak H$\alpha$
emitters. This may imply a link between the current accretion and
outflow rates: Hartmann et al.\ (1994) have shown that for TTS the bulk of the 
H$\alpha$ line emission comes from infalling, magnetospheric material and is 
not formed in the wind. Thus the  H$\alpha$ equivalent width, as we mentioned 
in the Introduction, is probably better 
regarded as a signature of accretion than of wind strength. This begs
the question, could some of the Group I stars without detected [OI] 
emission be undergoing episodes of reduced accretion activity? If this
is the case then their near-infrared colours could be due to the
presence of a optically thick reprocessing disk which would give 
H-K$\approx$~0.5.   

Finally we remark in connection with the other group for which we did not 
detect forbidden line emission, i.e.\ the Hillenbrand Group 3 stars, that a 
significant fraction may also not be genuine HAEBES. For example HD~37490 
(Omega Orionis) and  HD~76534 are almost certainly classical Be stars 
(Adelman 1992; Oudmaijer and Drew 1997). Similarly BD~+41~3731 was also 
rejected as a member of the HAEBES class (Th\'e et al.\ 1994) consistent 
with our observation of H$\alpha$ {\em in absorption} in this star.

In summary,  the relationship between the [OI] and H$\alpha$ equivalent 
widths with near-IR colours appears identical to that found for the
TTS.  It suggests, as with the TTS, that while some HAEBES are
surrounded by disks, others have a spectral energy distribution that
derives from a combination of a disk plus a halo. In a number of cases,
the Group III stars, either no disks are present or at best they are optically
thin, and it is precisely these stars that show very little sign of 
accretion and outflow activity. In contrast, their much redder
counterparts, the Group II stars show strong H$\alpha$ and [OI] emission.

\subsection{Luminosity relationships}

       Figs.\ 3a and 3b show good correlations between \LOI, \LHA\
and the infrared excess, \LIRE.  As with the TTS, this suggests 
a link between \massloss\ and \accrate\ in the HAEBES.
When combined with the results from Cabrit et al.\
(1990), in Fig.\ 5, the correlation between wind and accretion, in the
form of the \LOI\ vs.\ \LIRE\ plot, spans 5 orders of magnitude in \LIRE. 
It is difficult to imagine a purely stellar physical process that could
produce such a correlation especially when one considers that the 
sub-photospheric, photospheric and chromospheric conditions that occur 
in TTS are thought to be very different in HAEBES (for example, Catala 1989; 
Palla \& Stahler 1993). It is particularly interesting that the \LOI\ 
vs.\ \LIRE\ plot (as well as the \LHA\ vs.\ \LIRE\ plot which is not 
reproduced here) smoothly connects the two stellar groups. This clearly
suggests that disk plus halo models, which have met with much success in 
explaining many of the characteristics of CTTS, must be appropriate 
in a ``scaled-up'' version to their somewhat more massive counterparts.

	Other researchers' investigations of possible relationships
between \massloss\ and \accrate\ have produced similar results. 
Hamann \& Persson (1992) used the strongest line of the
calcium IR triplet, CaII$\lambda$8542, as a measure of \massloss\ in
both TTS and HAEBES and compared it with the infrared excess
calculated over a wavelength range 1 -- 25$\mu$m, using ground based
photometry and IRAS data.  The CaII$\lambda$8542 luminosities scale
with the infrared excess in the same manner as the emission line data
presented here. Hillenbrand et al.\ (1992) showed a correlation between
\LHA\ and \accrate\ for both TTS and HAEBES and found a result similar to
Fig.\ 5. Hillenbrand et al.\ (1992) used the monochromatic IR excess at 
3.5$\mu$m, taking into account the passive reprocessing of starlight by the 
disk material, as a measure of \accrate. As pointed out earlier on,
however, current models would imply that \LHA\ is not a measure of wind
strength and that any calculations of mass loss rates based on the flux
of the hydrogen lines, both optical and infrared, are ``meaningless''
(Calvet 1997). This criticism would also apply to the work of Nisini et al.\ 
(1995) who claimed to have directly measure \massloss\ in HAEBE stars using 
infrared HI recombination lines (Pa$\beta$, Br$\gamma$, Br$\alpha$ and 
Pf$\beta$). Moreover Nisini et al.\ (1995) suggested that TTS have higher
\massloss\ (based on TTS data from Giovanardi et al.\ 1991) than one would 
expect from extrapolating the \massloss\ vs.\ L$_{\rm bol}$ relationship
of the HAEBES to lower luminosities. Here L$_{\rm bol}$ is the bolometric
luminosity of the source and thus includes not only the stellar but the disk and
any halo component as well. Putting aside, for the moment, the claim by Calvet 
(1997) and others that the hydrogen emission lines in any event do not arise in
the wind, the suggestion that TTS are somehow more ``efficient'' at losing
mass than HAEBES is difficult to reconcile with our findings: we see a smooth 
transition in the relationship between wind strength and accretion across the 
TTS/HAEBES boundary (see Fig.\ 5). Moreover we also find a smooth relationship,
between  H$\alpha$ line strength and infrared excess. It appears, however, that 
the TTS mass-loss rates calculated by Giovanardi et al.\ (1991) are in any event
highly unreliable (Natta, private communication). In fact many of the TTS 
listed by Giovanardi et al.\ (1991) also had their mass loss rates estimated 
by Hartigan et al.\ (1995). These authors found values typically 10 times 
lower than Giovanardi et al.\ (1991), implying the ``efficiency'' of mass loss
is {\em not} higher amongst TTS than HAEBES and implying a much
smoother transition in \massloss\ vs.\ L$_{\rm bol}$ than found by 
Nisini et al.\ (1995; ibid Fig.\ 4). It is also worth remarking that a direct 
comparison of the \LIRE\ and \LSTAR\ values in Nisini et al.\ (1995) and ours 
reveals some distinct differences. It appears, however, that the calculated 
values of \LSTAR\ in Nisini et al.\ (1995) did not take account of the  
appropriate bolometric correction to M$_{\rm v}$. 

As can be seen from Figs.\ 4 \& 5, there is a tendency for \LOI, \LHA, and 
\LIRE\ to scale with the stellar luminosity, \LSTAR. In particular Fig.\ 4a
shows that there is a good correlation between \LIRE\ and \LSTAR\ and it is this
relationship  which, to some degree, masks whether \LOI\ and \LHA\ depend 
fundamentally on \LIRE\ or \LSTAR. The form of the dependence is
critical to understanding the origin of the forbidden line emission in HAEBES: 
Catala (1989) proposed a chromospheric origin for the winds from HAEBES that 
superficially might be in keeping with the dependence of \LOI, for example, on  
\LSTAR. According to that paradigm, the observed correlations with \LIRE\ 
would then be coincidental and due to the tendency for the total amount 
of circumstellar dust to scale with \LSTAR. Such an explanation, 
however, does not explain the tighter correlation we observe between 
\LOI\ and \LIRE\ than between \LOI\ and \LSTAR. Moreover, a chromospheric 
model seems incapable of explaining why a number of HAEBES show infrared excess
luminosities that are much greater than the underlying stars' bolometric
luminosities (Hamann \& Persson 1992).  Accretion must play a fundamental role 
as it is also impossible to imagine how such a situation can arise in a purely 
passive reprocessing disk or envelope. In fact, it may even be argued that the 
``reprocessing fraction'', particularly for the earliest stars, is lower amongst
HAEBES than TTS due to dust clearing (Hillenbrand et al.\ 1992; Hamann \& 
Persson 1992). 

A final point should be made in connection with the apparent dependence of 
wind properties on \LSTAR\ amongst HAEBES. In the case of the TTS, Cabrit 
et al.\ (1990) found a poor correlation of \LOI\ and \LHA\ with TTS 
photospheric luminosity in contrast to what is found for the HAEBES. The 
probable explanation for this effect is that in the TTS sample one is 
dealing with a large number of stars within a rather narrow luminosity (and 
mass) range. The sample of Cabrit et al.\ (1990) ranges in spectral type from 
M1 to G2 or a mass range of roughly 0.4 -- 1 M$_{\odot}$. The scatter in
such a sample of \LIRE, for a given narrow range in \LSTAR, is then  
large enough that the stronger dependence on \LIRE\ of \LOI , for example, 
is readily seen.  

It is interesting to compare the frequency of the occurrence of
molecular (CO) outflows with the near-infrared colour, or effectively as
one can see from Fig.\ 1, with Hillenbrand group. While none of our Group~III 
stars are observed to be accompanied by molecular outflows and only 2
out of our 21 Group I stars are, 6 out of our 9 Group II stars are found
to be associated with molecular outflows. Note that we used the molecular 
outflow database of Fukui et al.\ (1993) to check for associations. There thus 
appears to be a direct link between degree of embeddedness of the star and the 
presence or absence of large scale outflow phenomena.  The Group II 
stars' association with molecular outflows, and in some cases optical jets 
(Mundt \& Ray 1994),  may be due to their relatively early evolutionary state, 
as they disperse the dusty cocoons from which they formed. Strom et al.\
(1993) have suggested that the cessation of the outflow phase in TTS
is intimately linked to the circumstellar disk surrounding the star
becoming optically thin.  The Group III HAEBES may possess optically
thin circumstellar material and colder dust at large radii, a remnant
of an earlier phase, responsible for the longer wavelength IR excesses
observed in such stars.

\subsection{Correlations with the circumstellar mass}

In Fig.\ 6a we showed a plot of the luminosity of the [OI]$\lambda$6300 line 
against the calculated circumstellar mass. There is a clear correlation between 
the two quantities as is also the case for the infrared excess (\LIRE) and 
the circumstellar mass (see Fig. 6b and also Table 4). Obviously to an extent 
such correlations reflect the fact that the circumstellar mass tends to scale 
with the stellar luminosity \LSTAR\ and, given that  \LOI\ and \LIRE\ 
depend on \LSTAR, are to be expected. Nevertheless, Table 4 shows that 
the correlations of \LOI\ and \LIRE\ with M$_{{\rm cs}}$ are stronger than 
with \LSTAR\ suggesting some dependence of these quantities on M$_{{\rm cs}}$
alone. 
 
 The origin of the mm emission in HAEBES is not clear and this is partly due 
to the poor resolution of single dish mm measurements such as those reported 
here. As was stated in the 
Introduction, the SEDs of HAEBES have variously been modeled by extended dusty 
envelopes without disks (Berrilli et al.\ 1992; Di Francesco et al.\ 1994; 
Miroshnichenko et al.\ 1997) as well as by disk plus envelope combinations
(Hillenbrand et al.\ 1992; Natta 1993). In order to determine whether there is a
disk contribution to the millimeter emission, higher resolution mm-array 
observations are needed. To date, however, these have produced conflicting 
results (e.g., Di Francesco et al.\ 1997; Mannings \& Sargent 1997)  
In any event it is clear that if a major fraction of the circumstellar 
material inferred from our mm observations is confined to scales of 
typically a few arcseconds then, given its mass (see Table 5), one would 
expect much higher visual extinctions towards these sources than observed
where the material to be distributed spherically. We emphasize, however, that 
even if disks are present, as we believe they are, this does not exclude the 
possibility that some of the mm emission almost certainly arises from an
extended dusty halo (Di Francesco et al.\ 1994).

The correlations we observe in Fig.\ 6 do not tell us anything 
about the origin of the mm emission. They would be expected in the 
event that most of the mm emission came from an accretion disk but 
equally they might be anticipated if the mm emission largely derived 
from a dusty envelope that in turn fed a disk.

Finally we add a note of caution that, given our beam size, 
our mm flux measurements may include contributions from unrelated dust, 
and/or embedded companions, at large distances from the HAEBES. It is 
interesting to note, however, that our circumstellar mass estimates are 
close to those derived by Mannings \& Sargent (1997), from their much higher 
resolution observations, in at least the two stars common to both samples 
(i.e.\ AB~Aur and HD~163296). 

\section{Conclusions}

      The work of Cohen et al.\ (1989), Cabrit et al.\ (1990), Edwards et al.\ 
(1993), Hartigan et al.\ (1995) and others has demonstrated a strong link 
between accretion and mass-loss in the case of the CTTS. Using various
indicators of \accrate, the mass accretion rate, and \massloss, the
mass-loss rate, they have shown that the mass loss rate scales with  
the accretion rate and that the mass-loss most likely derives from an 
accretion-driven disk flow. The motivation for the present, and 
related (Corcoran \& Ray 1997a), work is to test whether the same basic
inflow/outflow process occurs amongst the HAEBES. The approach in this paper 
has been to test whether the same relationships seen in the TTS between 
indicators of mass loss and mass accretion carry over to their higher mass 
counterparts. Obviously if clear parallels can be drawn between the two groups 
this would not only bolster the idea that a common inflow/outflow mechanism 
operates amongst the TTS and HAEBES but indirectly, that disks, as in the 
case of CTTS, surround HAEBES (or at least those with substantial infrared/mm 
excesses). Certainly we expect that the disk paradigm can be extended to some 
intermediate mass young stars given the presence of dusty disks/rings around 
main sequence A-type stars like $\beta$~Pic (Burrows et al.\ 1997). The 
crucial question is rather how common are disks around HAEBES? The data 
and analysis we have presented here shows that in the case of the HAEBES:   

\begin {itemize} 

\item {Relative strengths of \OI\ and H$\alpha$ lines, as
determined from absolute equivalent widths, typically scale with 
reddened corrected near-IR colours, H-K and K-L for the Hillenbrand et al.\ 
(1992) Group I and II HAEBES. 
As discussed by Edwards et al.\ (1993) for the CTTS, these 
near-IR colours sample the temperature distribution of the inner system dust 
around the star. As the HAEBES near-IR colours increase above the values 
expected from a passive reprocessing or geometrically thin accretion disk, it 
is found that the relative wind strength increases in an exactly analogous 
manner to the CTTS. Since Hartigan et al.\ (1995) have shown that, at 
least in the case of the CTTS, the reddening corrected IR colours correlate 
well with the degree of veiling and hence indirectly with the relative accretion
rate, this strongly suggests that the same basic inflow/outflow process 
occurs in both CTTS and HAEBES. The idea that relative accretion is related 
to near-IR colour in HAEBES is further reinforced by the finding that 
for these stars K-L is correlated with \LIRE/\LSTAR\ where \LIRE\ is the 
infrared excess. \LIRE, if one ignores the effects of reprocessing, 
should be a rough measure of accretion.}  

\item {There are indications of a small population of Group I stars with 
no detected \OI\ emission and the levels of which are well below that 
expected on the basis of their near-IR colours. It seems unlikely that their 
infrared excesses can be explained by the presence of companions and the
absence of \OI\ emission may simply be due to variability as is certainly
true in a number of cases. Another possibility is that their disks are 
currently quiescent and merely reprocessing starlight.} 

\item {All Group II HAEBES have near-IR colours that are much redder than
those of a passive reprocessing or geometrically thin accretion
disk. As noted by Hillenbrand et al.\ (1992) their overall spectral
energy distribution is best explained in terms of a star+disk+envelope
model. These are found to be the the stars with the strongest evidence for 
accretion. We note that the Group II stars are, proportionately, by far the most
likely sources of extended optical jets amongst all HAEBES (Corcoran \& Ray 
1997a).}

\item {No forbidden line emission was detected from any of our Hillenbrand et 
al.\ (1992) Group III stars in keeping with the small values of their 
near-infrared colours (i.e.\ close to that expected from a main sequence star of
the same spectral type). Moreover the equivalent width of their H$\alpha$ 
emission is $\leq$ 15\AA\ in all cases. We propose that their lack of forbidden 
line activity and, in particular such an H$\alpha$ equivalent width limit, may 
serve as a useful criteria  when searching for members of the Group III HAEBES 
sub-class. That said, we should warn the reader that the 
weak-line HAEBES, ``WHAEBES'', defined in this way are not exactly identical to 
the Group III stars as defined by Hillenbrand et al.\ (1992). The latter are 
characterised in terms of the near-infrared slope of their SEDs whereas the 
WHAEBES will also include those Group~I/II stars with low line activity (e.g.\ 
BF~Ori, MWC~137, HD~150193), the status of which is still unclear. While it 
seems some Group~III stars are not pre-main sequence stars, nevertheless it 
appears likely that a number represent the HAEBES analogues of the weak-line T 
Tauri stars. Interestingly Group III stars are predominantly early B-type stars.
Such stars evolve rapidly and will therefore not have wandered far from their 
formation sites. On the other hand the Group I and II stars are predominately 
of later spectral type. This begs the question where are the Group III late-type
HAEBES? Such stars should be searched for, perhaps many of them lie well 
beyond the boundaries of their parent molecular cloud. }

\item {Using the luminosity of the [OI]$\lambda$6300 emission line as an 
indirect measure of \massloss\ and the IR excess luminosity over 
0.7--10.2$\mu$m as a measure of \accrate, a strong correlation is found 
between accretion and mass-loss. While the [OI]$\lambda$6300 line 
luminosity also tends to increase with stellar luminosity, as one might 
expect, the correlation with IR excess luminosity {\em is stronger} 
implying that accretion rate is an important factor in determining the 
strength of the [OI] emission.}

\item {Plotted to the same axes as our HAEBES data, the \LOI\ vs.\
\LIRE\ of the sample of 36 TTS from Cabrit et al.\ (1990) extends the
correlation between the two variables over 5 orders of magnitude in
luminosity. The smooth progression from PMS stars of less than a solar
mass all the way to B-type stars with masses $\sim$ 10M$_{\odot}$ or
more provides convincing evidence that the origin of the \massloss\
$\propto$ \accrate\ relationship is the same process in both groups of
stars. Given the compelling evidence in the case of CTTS that matter
is accreted via a disk, one must infer that accretion disks also
surround many HAEBES.}

\item {Millimeter continuum observations of HAEBES reveal
circumstellar masses typically in the range 0.02--1.0 M$_{\odot}$, assuming a
gas to dust ratio of 100:1 and optically thin emission. It is likely in a number
of cases that these masses are somewhat overestimated due to our large beam 
size (30$''$). If a significant proportion of the mm flux comes from a 
compact region on scales less than a few arcseconds then the 
masses inferred would imply much higher optical extinction towards these stars 
than observed unless the dust is {\em not} distributed spherically.
Recent observations support the idea that the emission is compact (Mannings 
\& Sargent 1997) and this again suggests that disks may be present around many 
HAEBES.  The amount of circumstellar mass, M$_{{\rm cs}}$, present is 
slightly better correlated with the luminosity of the forbidden line 
emission than with the photospheric luminosity of the star. Again this 
indicates that a disk or a disk+halo combination may be fueling HAEBES
with forbidden line emission.}

\end {itemize}

\acknowledgements
   MC would like to acknowledge funding from Fobairt, the Irish
   Science and Technology Agency. The authors also thank D.\ Corcoran
   and A.\ Moorhouse for their observations on the INT and
   ESO/MPI~2.2m Telescope respectively and to A.\ Sargent, S.V.W.\
   Beckwith and C.\ Koresko for their help with the CSO observations. We
   would also like to thank very sincerely the referee, S.\ Cabrit,
   for her patience and very helpful comments. Finally, we would like to
   express our gratitude to the staff of the La Palma Observatory for
   their assistance.  The Isaac Newton Telescope on the island of La Palma
   is operated by the Royal Greenwich Observatory at the Spanish
   Observatorio del Roque de los Muchachos of the Instituto de
   Astrofisica de Canarias. This research has made use of the Simbad
   database, operated at CDS, Strasbourg, France.
%

\vspace{1.0cm}

\noindent{\bf References}

\rf {Adams, F.C., Lada, C.J., \& Shu, F.H., 1987, \APJ {312} {788}}
\rf {Adelman, S.J., 1992, \PASP {104} {392}}
\rf {Beckwith, S.V.W., Sargent, A.I., Chini, R.S., \& G\"usten, R., 1990, \AJ
{99} {924}}
\rf {Beckwith, S.V.W., \& Sargent, A.I., 1991, \APJ {381} {250}}
\rf {Berrilli, F., Corciulo, G., Ingrosso, G., Lorenzetti, D., Nisini,
B., \& Strafella, F., 1992, \APJ {398}{254}}
\rf {Bertout, C., Basri, G., \& Bouvier, J., 1988, \APJ {330} {350}}
\rf {B\"ohm, T., \& Catala, C., 1994, \AAP {290} {167}}
\rf {B\"ohm, T., \& Catala, C., 1995, \AAP {301} {155}}
\rf {Burrows, C., et al., 1997, ApJ, in press}
\rf {Cabrit, S., Edwards, S., Strom, S. E. \& Strom, K. M., 1990, \APJ {354} {687}}
\rf {Calvet, N., 1997, in: Herbig-Haro Outflows and the Birth of Low
Mass Stars, IAU Symposium No.\ 182, eds.\ B.\ Reipurth and C.\ Bertout, 
(Kluwer Academic Publishers), p.\ 417}
\rf {Catala, C., 1989, in: ESO workshop on Low Mass Star Formation and Pre-Main Sequence Objects ed.\ B.\ Reipurth (Garching: European Southern Obs.) p.\ 471}
\rf {Cayrel, R., 1990, in: Proceedings of the Workshop on Physical Processes in Fragmentation and Star Formation, (Dordrecht,  Kluewer Academic Publishers), p.\ 343}
\rf {Cohen, M., Emerson, J., \& Beichman, C., 1989, \APJ {339} {455}}
\rf {Corcoran, M., \& Ray, T. P., 1996, in: Proceedings of Disks and Outflows 
Around Young Stars, eds.\ Beckwith, S.V.W., Natta, A., \& Staude, J., 
Lecture Notes in Physics Series, (Heidelberg: Springer Verlag), p.\ 276}
\rf {Corcoran, M., \& Ray, T. P., 1997a, \AAP {321} {189}}
\rf {Corcoran, M., \& Ray, T. P., 1997b, in preparation}
\rf {Di Francesco, J., Evans, N. J., Harvey, P. M., Mundy, L. G., \& Butner, H. M., 1994, \APJ {432} {710}}
\rf {Di Francesco, J., Evans, N. J., Harvey, P. M., Mundy, L. G., 
Guilloteau, S., \& Chandler, C.J., 1997, \APJ {482} {433}
\rf {Edwards, S.E., Hartigan, P., Ghandour, L.,  Andrulis, C., 1994, \AJ {108} {1056}}
\rf {Edwards, S.E., Ray, T.P. \& Mundt, R., 1993, in: Protostars and Planets III, eds.\ E.H.\ Levy \& J.I.\ Lunine,  (Tucson: University of Arizona), p.\ 567}
\rf {Finkenzeller, U., \& Mundt, R., 1984, \APJS {55} {109}}
\rf {Fukui, Y., Iwata, T., Mizuno, A., Bally, J., \& Lane, P., 1993, in: Protostars and Planets III, eds.\ E.H.\ Levy \& J.I.\ Lunine,  (Tucson: University of Arizona), p.\ 603}
\rf {Ghandour, L., Strom, S., Edwards, S., \& Hillenbrand, L., 1994,
in: The Nature and Evolutionary Status of Herbig Ae/Be Stars, eds.\
Th\'e, P.S., P\'erez, M.R., \& van den Heuvel, P.J., (Astronomical
Society of the Pacific Conference Series Vol.\ 62), 223}
\rf {Gilliland, R.L., 1986, \APJ {300} {339}}
\rf {Giovanardi, C., Gennari, S., Natta, A., \& Stanga, R., 1991, \APJ {367} {173}}
\rf {Grady, C.A., P\'erez, M.R., Talavera, A., Th\'e, P.S., de Winter, D., Grinin, V.P., \& Calvet, N., 1993, \BAAS {183}, {41.09}}
\rf {Hamann, F., \& Persson, S.E. 1992, \APJ {394} {628}}
\rf {Hartigan, P., Edwards, S.E., \& Ghandour, L., 1995 \APJ {452} {736}}
\rf {Hartmann, L., Hewett, R., \& Calvet, N., 1994, \APJ {426} {669}}
\rf {Hartmann, L., Kenyon, S.J., \& Calvet, N., 1993, \APJ {408} {219}}
\rf {Henning, Th., Launhardt, R., Steinacker, J. \& Thamm, E., 1994, \AAP {291} {546}}
\rf {Herbig, G.H., 1960, \APJS {4} {337}}
\rf {Hillenbrand, L.A., Strom, S.E., Vrba, F.J.\ \& Keene, J.\ 1992, \APJ {397} {613}}
\rf {Imhoff, C.L., 1994, in: The Nature and Evolutionary Status of Herbig Ae/Be Stars, eds.\ Th\'e, P.S., P\'erez, M.R., \& van den Heuvel, P.J., (Astronomical Society of the Pacific Conference Series Vol.\ 62), p.\ 107}
\rf {Isobe, T., LaValley, M. \& Feigelson, E. D. 1986, \APJ {306} {490}}
\rf {Jain, S.K. \& Bhatt, H.C., 1995, \AAS {111} {39}}
\rf {Kenyon S.J., \& Hartmann, L., 1987, \APJ {323} {714}}
\rf {Kenyon S.J., \& Hartmann, L., 1990, \APJ {349} {197}}
\rf {Leinert, C., Richichi, A., \& Haas, M., 1997, \AAP {318} {472}}
\rf {Leverault, R.M., 1988, \APJ {330}{897}}
\rf {Mannings, V., 1994, \MN {271} {587}}
\rf {Mannings, V., \& Sargent, A.I., 1997, ApJ, in press} 
\rf {Men'shchikov, A.B., \& Henning, Th., 1997, \AAP {318} {879}}
\rf {Miroshnichenko, A., Ivezi\'c, C.Z., \& Elitzur, M., 1997, \APJ {475} {L41}}
\rf {Mundt, R., \& Ray, T.P., 1994, in: The Nature and Evolutionary Status of 
Herbig Ae/Be Stars, eds.\ Th\'e, P.S., P\'erez, M.R., \& van den Heuvel, P.J., 
(Astronomical Society of the Pacific Conference Series Vol.\ 62), p.\ 237}
\rf {Natta, A., 1993, \APJ {412} {761}}
\rf {Nisini, B., Milillo, A., Saraceno, P., \& Vitali, F., 1995, \AAP {302} {169}}
\rf {Oudmaijer, R.D., Drew, J.E., 1997, \AAP {318} {198}}
\rf {Palla, F., \& Prusti, T., 1993, \AAP {272} {249}}
\rf {Palla, F., \& Stahler, S.W., 1993, \APJ {418} {414}}
\rf {P\'erez, M.R., Imhoff, C.L., \& Th\'e, P.S., 1992, \BAAS {23} {1374}}
\rf {Pirzkal, N., Spillar, E.J., \& Dyck, H.M., 1997, \APJ {481} {392}}
\rf {Pollack, J.B., Hollenbach, D., Beckwith, S., Simonelli, D.P., Roush, T., \& Fong, W., 1994, \APJ {421} {615}}
\rf {Scheffler, H., \& Els\"asser, H., 1982, Physics of the Galaxy and 
Interstellar Matter, trans.\ Armstrong A.H. (Springer-Verlag), p.\ 152}
\rf {Skinner, S.L., Brown, A., \& Stewart, R.T., 1993, \APJS {87}{217}}
\rf {Sorelli, C., Grinin, V.P., Natta, A.\ 1996, \AAP {309}{155}}
\rf {Strom, S.E., Edwards, S., \& Skrutskie, M.F., 1993, in: Protostars and Planets III, eds.\ E.H.\ Levy \& J.I.\ Lunine,  (Tucson: University of Arizona), p.\ 837}
\rf {Th\'e, P.S., Cuypers, H., \& Tijn A Djie, H.R.E., 1985, \AAP {149}
{429}}
\rf {Th\'e, P.S., De~Winter, D., \& Perez, M.R., 1994, \AAS {104}
{315}}
\rf {Vigneron, C., Mangeney, A., Catala, C., Schatzman, E., 1990, Solar
Phys.\ 128, 287}
\rf {Zombeck, M.V., 1990, Handbook of Space Astronomy and Astrophysics, 
(Cambridge University Press)}

\end{document}